\newcommand{\preprintDate}{27 October 2023}

\documentclass{jfm-arxiv}

\usepackage[T1]{fontenc}
\usepackage[utf8]{inputenc}
\usepackage[UKenglish]{babel}
\usepackage[style=british]{csquotes}
\usepackage{microtype}

\usepackage{graphicx}
\usepackage{newtxtext}
\usepackage{newtxmath}

\usepackage[breaklinks]{hyperref}
\hypersetup{
    colorlinks = true,
    urlcolor   = blue,
    citecolor  = black,
}
\usepackage[section]{placeins}

\usepackage[backend=biber,natbib=true,style=authoryear-comp,sorting=nyvt,sortlocale=en-GB,maxbibnames=999,maxcitenames=2,mincitenames=1,date=year,labeldate=year,alldates=year]{biblatex} % modern citation
\bibliography{ms}
\renewbibmacro{in:}{}

\usepackage{xcolor}
\usepackage{overpic}
\usepackage{mathtools}
\usepackage{ifthen}
\usepackage{gensymb}
\usepackage[normalem]{ulem}
\usepackage{enumitem}

\graphicspath{{figures/}}

\makeatletter
\g@addto@macro\@floatboxreset{\centering}
\makeatother
\usepackage{float}

\newcommand{\secref}[1]{\S\ref{sec/#1}}
\newcommand{\figref}[1]{figure~\ref{fig/#1}}

\newcommand{\Figref}[1]{Figure~\ref{fig/#1}}

\newcommand{\ppf}{ppf}
\newcommand{\pcf}{pcf}

\newcommand*\samethanks[1][\value{footnote}]{\footnotemark[#1]}

\title{Dynamics and proliferation of turbulent stripes in plane-Poiseuille and plane-Couette flows}
\author{E.\ Marensi\aff{1,2}\corresp{\email{e.marensi@sheffield.ac.uk}}
    \thanks{These authors contributed equally to this work.},
    G.\ Yalnız\aff{1}\samethanks{}
    \and B.\ Hof\aff{1}}
\affiliation{\aff{1}Institute of Science and Technology Austria (ISTA), 
                Am Campus 1, 3400 Klosterneuburg, Austria
            \aff{2}The University of Sheffield, Department of Mechanical Engineering,
            Mappin Street, S1 3JD Sheffield, United Kingdom}

\pubyear{2023}
\date{\preprintDate}
\begin{document}
\maketitle

\begin{abstract}
The first long-lived turbulent structures observable in planar shear flows take the form of localized stripes, inclined with respect to the mean flow direction. The dynamics of these stripes are central to transition, and recent studies proposed an analogy to directed percolation where the stripes' proliferation is ultimately responsible for turbulence to become sustained. In the present study we focus on the internal stripe dynamics as well as on the eventual stripe expansion, and we compare the underlying mechanisms in pressure and shear driven planar flows, respectively plane-Poiseuille and plane-Couette flow. Despite the similarities of the overall laminar-turbulence patterns, the stripe proliferation processes in the two cases are fundamentally different. Starting from the growth and sustenance of individual stripes, we find that in plane-Couette flow new streaks are created stochastically throughout the stripe whereas in plane-Poiseuille flow streak creation is deterministic and occurs locally at the downstream tip. Because of the up/downstream symmetry, Couette stripes, in contrast to Poiseuille stripes, have two weak and two strong laminar turbulent interfaces. These differences in symmetry as well as in internal growth give rise to two fundamentally different stripe splitting mechanisms. In plane-Poiseuille flow splitting is connected to the elongational growth of the original stripe, and it results from a break-off / shedding of the stripe's tail. In plane-Couette flow splitting follows from a broadening of the original stripe and a division along the stripe into two slimmer stripes.
\end{abstract}

\section{Introduction}\label{sec/intro}
In wall bounded shear flows turbulence often arises despite the linear stability of the laminar state \citep{drazin2004hydrodynamic}. Under such circumstances, depending on initial conditions and on the presence of finite amplitude perturbations, flows can be either laminar or turbulent. Moreover, at moderate Reynolds numbers turbulence cannot be excited uniformly throughout space but it remains restricted to localized domains \citep{reynolds1883xxix, wygnanski1973transition, carlson1982flow,alavyoon1986turbulent, daviaud1992subcritical, tillmark1992experiments,prigent2002largescale, tsukahara2005dns}, which require energy input from adjacent laminar regions. As a result flows are spatio-temporally intermittent and consist of alternating laminar and turbulent regions. For pipe flow where laminar-turbulent patterns are essentially one dimensional, it has been shown that the spacing between turbulent patches (puffs) is related to an energy depletion of the flow and the necessity of the velocity profile to recover its more energetic parabolic shape \citep{vandoorne2008flow,hof2010eliminating,samanta2011experimental}. Indeed, in the experiments of \cite{hof2010eliminating} it was shown that if two puffs were triggered too close to each other, the downstream puff would collapse due to the plug-shaped energy-depleted flow exiting the upstream puff. For flows extended in two dimensions such as plane-Couette flow (\pcf{}) and plane-Poiseuille flow (\ppf{}), the situation is more complex and turbulence tends to appear in the form of stripes that are inclined with respect to the wall movement or bulk flow direction \citep{coles1965transition,prigent2002largescale,tsukahara2005dns, tuckerman2020patterns}. In analogy to pipe flow it has been suggested by \cite{samanta2011experimental} that also in this two-dimensional setting the finite width of stripes is related to the energy depletion of the mean velocity profile and therefore the apparent wavelength of stripe patterns is set by local interactions and not by some global instability \citep{prigent2002largescale}. Recently, the mechanism of wavelength selection in Couette flow has been associated by \cite{gome2022patterns} to an energy maximization principle of the large-scale circulatory flow surrounding stripes.

% Isolated stripe, coordinate systems
\begin{figure}
    \begin{overpic}[width=0.733\linewidth]{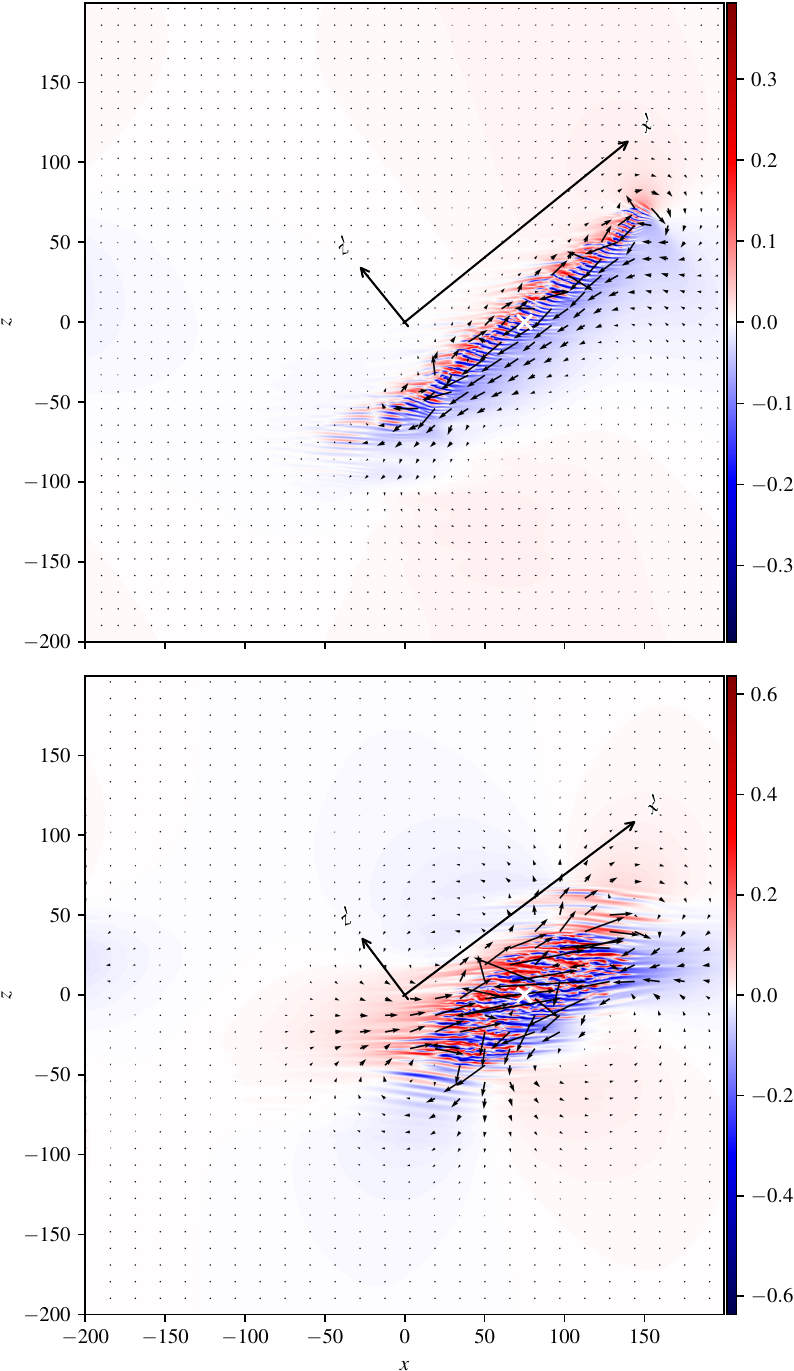}
        \put (0, 100) {(a)}
        \put (0, 50)  {(b)}
    \end{overpic}
    \caption{Streamwise velocity fluctuations $u_x$ (colours) and in-plane velocity fluctuations $\{u_x, u_z\}$ (arrows) of (a) plane-Poiseuille flow ($\Rey^P=660$) and (b) plane-Couette flow ($\Rey^C=335$). Colours are the values at the wall-normal planes \(y=0.46\) for \ppf{} and \(y=0\) for \pcf{}, respectively, where the mean velocity profiles intersect the laminar profiles. Displayed arrows are proportional, with the same scale in (a,b), to in-plane velocity fluctuations averaged in the wall-normal direction. Tilted coordinates $x'$-$z'$ correspond to the stripe-parallel and stripe-perpendicular directions. Crosses ($\times$) mark the centre of `mass', where mass is taken to be $u_x^2$. (We take an average of the grid point locations weighted with $u_x^2$, respecting the periodicity of $x$ and $z$, see \secref{stripe-cm} for details.).}
    \label{fig/isolated-stripe}
\end{figure}

In the present study we focus on isolated stripes (see \figref{isolated-stripe}) which already appear at somewhat lower Reynolds numbers. Just like individual puffs in pipe flow, they tend to live for long times but ultimately decay, i.e.\ their lifetimes remain finite \citep{bottin1998discontinuous,borreroecheverry2010transient,shi2013scale, shimizu2019exponential,avila2021second}. For turbulence to become sustained a proliferation process is required that can balance the decay of individual patches. This competition between decay and proliferation \citep{avila2011onset} gives rise to a phase transition separating flows that eventually relaminarize from those where spatio-temporal intermittency persists indefinitely. As shown in recent studies this onset of sustained turbulence in some cases (e.g.\ Couette and Waleffe flow) falls into the directed percolation universality class \citep{lemoult2016directed,shih2015ecological,chantry2017universal,klotz2022phase,hof2023directed}. Conversely, in extended Poiseuille flow domains, recent experiments \citep{mukund2021aging} have suggested that the transition does not fall into the directed percolation universality class, an observation that has been related to a hydrodynamic instability happening at the downstream tip of fully localized Poiseuille stripes \citep{xiao2019growth,kanazawa2018thesis}. Nevertheless, it is still possible to determine a critical point for the onset of sustained turbulence by balancing the decay and growth mechanisms.

In flows extended in one dimension, such as pipe flow, the proliferation driving the transition \citep{moxey2010distinct} occurs via puff splitting. In this process \citep{wygnanski1975on} vortices are shed from an existing puff and if they can separate sufficiently far from the original puff, a new puff can grow. For planar Couette and Poiseuille flows growth can occur in two dimensions, i.e.\ stripes can split but they can also elongate. In particular for fully localized stripes the nature of these processes is not well understood, and it remains unclear if the stripe growth and splitting processes are the same in these two geometries. 

In the following, after describing the simulation details and coordinate systems (\secref{numerics}), we will show that the internal streak creation (\secref{growth}) and transport processes (\secref{transport}) of isolated stripes are fundamentally different in Poiseuille and Couette flows. These differences, together with the different flow symmetries, will be used to guide our analysis of laminar-turbulent interfaces (\secref{interfaces}) and to explain the different splitting mechanisms (\secref{splitting}) in the two flows. Some concluding remarks will be given in \secref{conclusions}.

\section{Numerical details and coordinates}\label{sec/numerics}
We perform direct numerical simulations of shear-driven and pressure-driven channel flows in extended domains of streamwise and spanwise lengths $L_x=L_z=400 h$, where $h$ is the half-gap width of the channel ($L_y=2h$, defining the volume $V=L_x L_y L_z$, and the average $\langle \square \rangle = V^{-1} \int\mathrm{d}x\mathrm{d}y\mathrm{d}z\, (\square)$ for any quantity $\square$). Here, $\{x, y,z\}$ indicate the streamwise, wall-normal and spanwise directions, and $\{\mathbf{e}_x, \mathbf{e}_y,\mathbf{e}_z\}$ are the corresponding unit vectors. The wall speed $U_w$ in plane-Couette flow and the centreline velocity $U_\mathrm{cl}$ in plane-Poiseuille flow are used as velocity scales, and $h$ is used as the length scale in both systems. The Reynolds numbers in the two flow geometries are thus defined as $\Rey^{C}=U_w h/\nu$ and $\Rey^{P}=U_\mathrm{cl} h/\nu$, respectively, where $\nu$ is the kinematic viscosity of the fluid. The total velocity and pressure are decomposed as $\mathbf{u}_{\mathrm{tot}}(\mathbf{x}, t) = \mathbf{U}_{\mathrm{lam}}(y) + \mathbf{u}(\mathbf{x}, t)$ and $p_{\mathrm{tot}}(\mathbf{x}, t) = \Pi(t)x + p (\mathbf{x}, t)$ where $\mathbf{u} = \{u_x, u_y, u_z\}$ is the deviation from the laminar velocity and $\Pi(t) = \langle \partial p_{\mathrm{tot}}/ \partial x \rangle$ is the mean pressure gradient. The dimensionless laminar flow $\mathbf{U}_{\mathrm{lam}} = U(y)\mathbf{e}_x$ is given by $U := U^{C} = y$ in plane-Couette flow and $U := U^{P} = 1-y^2$ in plane-Poiseuille flow. The latter has a non-zero bulk velocity in the streamwise direction, $U_{\mathrm{bulk}}^P = \langle 1-y^2 \rangle =2/3$. We run simulations of plane-Poiseuille flow such that the bulk velocity is kept constant at all times, $U_{\mathrm{bulk}}(t) = \langle u_\mathrm{tot} \rangle = U_{\mathrm{bulk}}^P$, and allow the mean pressure gradient $\Pi(t)$ to vary instead. For plane-Couette flow $\Pi(t)$ is fixed to zero. To compare time intervals between the two systems, we will refer to advective time units, which are $t_\mathrm{advective} =L_y/U_\mathrm{bulk}=3$ for plane-Poiseuille flow, and $t_\mathrm{advective} = L_y/U_w=2$ [equivalent to $1/\int_0^1 \mathrm{d}y\, u_\mathrm{lam}(y)$] for Couette flow.
 
The evolution equations are the Navier--Stokes and continuity equations for an incompressible Newtonian fluid, and are solved numerically using a hybrid solver (psuedo-spectral in the homogenous directions, finite differences in the wall-normal direction), adapted from Openpipeflow \citep{willis2017openpipeflow} for planar geometries. The spatial resolution is $(N_x, N_y, N_z)=(1024, 49, 1536)$ and $(N_x, N_y, N_z)=(1536, 65, 1536)$, for plane-Couette and plane-Poiseuille flows respectively, where $N_x$ and $N_z$ are the number of Fourier modes in the streamwise and spanwise directions (prior to a $3/2$ zero padding of the modes for dealiasing) and $N_y$ is the number of grid points (extrema of Chebyshev polynomials defined in $y \in [-1,1]$) in the wall-normal direction. The resolution is chosen so that it ensures a drop-off of at least three orders of magnitude in the amplitude spectra. The resulting streamwise and spanwise grid spacings, in units of $h$, are $\Delta_x=0.26$ and $\Delta_z=0.17$ for \pcf{} and $\Delta_x=\Delta_z=0.17$ for \ppf{}. A second-order predictor-corrector scheme is used for temporal discretization with a fixed timestep of $0.01$ which gives a Courant--Friedrichs--Lewy (CFL) number no larger than $0.2$.

\subsection{Tilted coordinates}\label{sec/tiltedcoords}
Given that stripes are tilted with respect to the streamwise direction (see \figref{isolated-stripe}), we will often work with coordinates that align better with the stripes' orientations. To this end, we define the $\theta$-rotated (counterclockwise) versions of the $x-z$ axes,
 \begin{equation}
   \begin{aligned}
     \label{totilted}
     x' &= \cos\theta\, x + \sin\theta\, z\,,\\
     z' &= -\sin\theta\, x + \cos\theta\, z\,,
   \end{aligned}
 \end{equation}
along with the corresponding unit vectors
 \begin{equation}
   \begin{aligned}
     \mathbf{e}_{x'} &= \cos\theta\, \mathbf{e}_x + \sin\theta\, \mathbf{e}_z\,,\\
     \mathbf{e}_{z'} &= -\sin\theta\, \mathbf{e}_x + \cos\theta\, \mathbf{e}_z\,.
   \end{aligned}
 \end{equation}
We call $x'$ as the `stripe-parallel' and $z'$ as the `stripe-perpendicular' coordinate, respectively. These coordinates with particular choices of $\theta$ are displayed in \figref{isolated-stripe}. For \ppf{} we set $\theta:=\theta^{P}=39^\degree$, and for \pcf{} we set $\theta:=\theta^{C}=37^\degree$. These are the maxima of the angles we have observed within stripes' (of $\Rey^{C}=335$ and $\Rey^{P}=660$) time evolution: instantaneous stripe angles (see \secref{stripe-theta} for the definition and \figref{inst-angles} for the time series) were at most $3^{\degree}$ less and the averages $1^{\degree}$ less than these fixed angles. We would like to note that these coordinates do not follow the stripes, $\theta$ is fixed, and $(x',z')=(0,0)$ always coincides with $(x,z)=(0,0)$.
 
We handle the rotation of the fields by $\theta$, as post-processing, with a function (\texttt{ndimage.rotate}) from SciPy \citep{SciPy}, a Python library, which handles the padding of the original domain with periodic copies prior to the rotation as well as the rotation itself. In the rest of this article, we will refer to $u_{x'}=\mathbf{e}_{x'}\cdot\mathbf{u}$ as the `stripe-parallel' and $u_{z'}=\mathbf{e}_{z'}\cdot\mathbf{u}$ as the `stripe-perpendicular' velocity fluctuations, respectively.

\section{Results}\label{sec/results}
In order to obtain isolated stripes in both plane-Couette and plane-Poiseuille flow we start from simulations of featureless turbulence and gradually decrease the Reynolds number until a single stripe survives which remains approximately stable (neither growing nor decaying). This condition is met close to the critical point where stripes first become sustained in the respective flow. For \pcf{} this corresponds to $\Rey^C\approx 325$ \citep{bottin1998discontinuous,duguet2010formation} and for \ppf{} $\Rey^P \approx 650$ \citep{mukund2021aging}. Such stripes are then used as initial conditions for simulations slightly above the respective critical points, specifically at $\Rey^C=335$ and $\Rey^P=660$, at which individual stripes slowly grow. Simulations at these Reynolds numbers will be used in \S\ref{sec/growth}--\ref{sec/interfaces} to study the growth and sustenance of individual stripes as well as the stripe internal dynamics and laminar-turbulent interfaces. Simulations of stripe splitting, which will be analysed in section \secref{splitting}, are performed at higher Reynolds numbers, namely at $\Rey^C=350$ and $\Rey^P=750$.

\subsection{Streak generation}\label{sec/streakgen}
\label{sec/growth}
Isolated stripes in \ppf{} and \pcf{} are displayed in \figref{isolated-stripe} for $\Rey^P=660$ and $\Rey^C=335$, respectively. Their time evolution is shown in the Supplementary Movies 1 and 2. The stripes are visualized at wall-normal planes where the mean (streamwise and spanwise averaged) velocity profiles intersect the laminar profiles, with colours corresponding to the streamwise velocity fluctuations, and arrows corresponding to the in-plane velocity fluctuations (averaged in the wall-normal direction). The latter highlight the presence of a secondary flow which develops around fully localized turbulent structures \citep{lundbladh1991direct,lemoult2013turbulent, duguet2013oblique,couliou2015largescale,klotz2021experimental}. At the chosen wall-normal planes, positive and negative streamwise velocity fluctuations are (roughly) equally frequent so that the alternation of high-speed (red regions) and low-speed (blue regions) streaks is apparent in both geometries. Based on visual inspection, streak and vortex patterns in the stripes' interior appear to be more regular for Poiseuille stripes than for Couette stripes. This difference is also clearly visible in the Supplementary Movies 1 and 2.
 
\begin{figure}
    \begin{overpic}[width=0.99\linewidth]{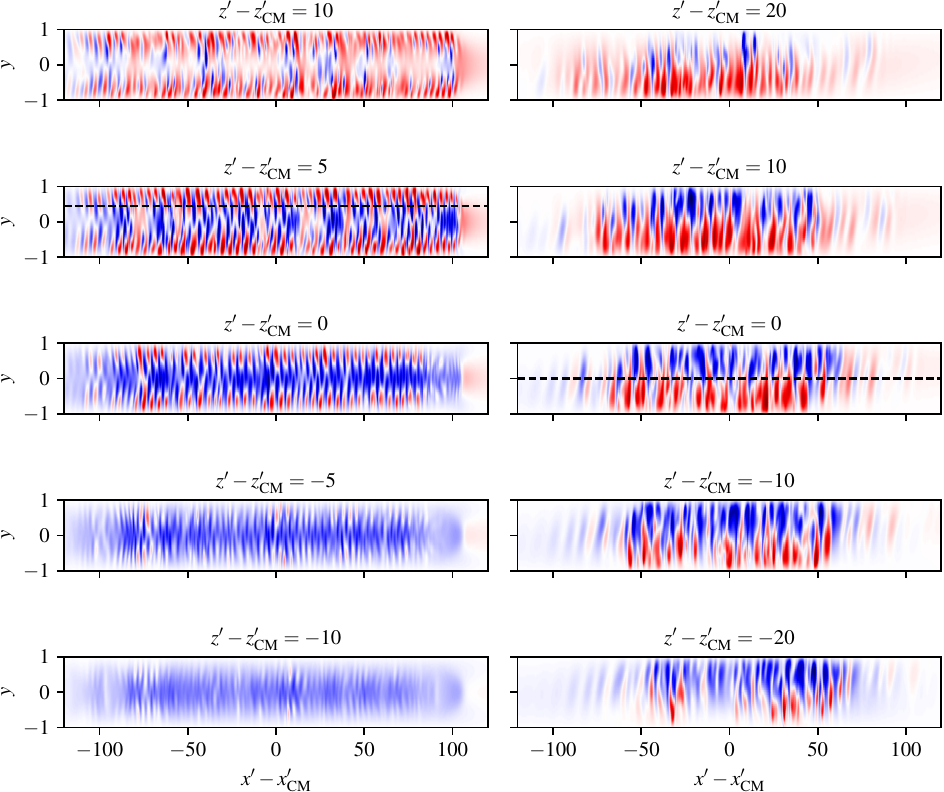}
        \put (0, 82)  {(a)}
        \put (0, 66)  {(b)}
        \put (0, 50)  {(c)}
        \put (0, 33)  {(d)}
        \put (0, 16)   {(e)}

        \put (52, 82)  {(f)}
        \put (52, 66)  {(g)}
        \put (52, 50)  {(h)}
        \put (52, 33)  {(i)}
        \put (52, 16)   {(j)}
    \end{overpic}
    \caption{Streamwise velocity fluctuations $u_x$ of (a--e) \ppf{} and (f--j) \pcf{} at various stripe-perpendicular ($z'$) locations. The centres of mass ($x'_\mathrm{CM}, z'_\mathrm{CM}$) were marked in \figref{isolated-stripe}. Dashed lines in (b) and (h) indicate the wall-normal planes at which the laminar and mean turbulent profile intersect. They are marked in the panels corresponding to the stripe-perpendicular distances to the centres of mass (noted on top of each panel) which will be selected for the spacetime figures \ref{fig/spacetime} and \ref{fig/spacetime2} to follow.}
    \label{fig/isolated-stripe-cuts}
\end{figure}
 
% Spacetime plots
\begin{figure}
    \begin{overpic}[width=0.99\linewidth]{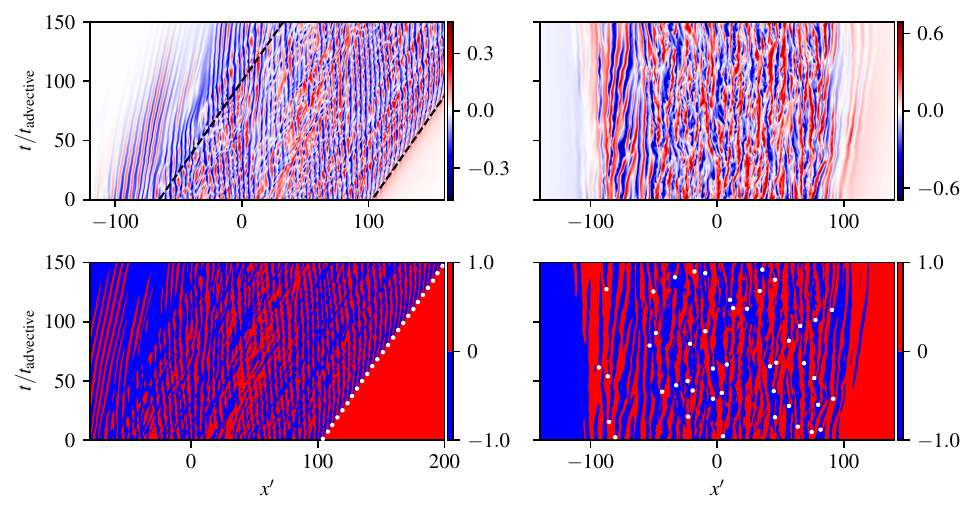}
        \put (0, 51)  {(a)}
        \put (51, 51)  {(c)}
        \put (0, 26)  {(b)}
        \put (51, 26)  {(d)}
    \end{overpic}
    \caption{Spacetime plots of streamwise velocity fluctuations in the bulk frames of reference of (a,b) \ppf{} and (c,d) \pcf{}. Colours are the values of (a) $u_x(t,x' + tU_\mathrm{bulk}\cos\theta^P)|_{y=0.46,\, z'=z'_\mathrm{CM}(t)+5}$ for \ppf{} and (c) $u_x(t,x')|_{y=0,\, z'=z'_\mathrm{CM}(t)}$ for \pcf{} (see dashed lines in panels (b) and (h) of \figref{isolated-stripe-cuts}). Panels (b,d) are binary versions of (a,c): anywhere with positive/negative streamwise velocity fluctuation is shown red/blue and newly created streaks are marked with white dots. For visualizations reasons, only the markings of positive (red) new streaks are shown for \ppf{}. Dashed black line downstream in (a) is fitted to the streak creation events marked in (b), the line upstream with the same slope is put as a guide.}
    \label{fig/spacetime}
\end{figure}
 
For stripes to become sustained and grow, streaks and vortices must be (re)generated and as we will see below, the growth of individual stripes is a requirement for the creation of new stripes (i.e.\ for splittings). First we will focus on individual stripes and the internal creation of new streaks. To gain some insight into this mechanism, we visualize (\figref{isolated-stripe-cuts}) streaks in stripe-parallel ($x'-y$) planes. Planes are selected with respect to the stripe's centre of mass (see \secref{stripe-cm} for details) at $z'=z'_\mathrm{CM}(t)+\Delta$, where $\Delta=0,\pm 5, \pm 10$ for \ppf{} and $\Delta=0, \pm 10, \pm 20$ for \pcf{} (the difference in $\Delta$ is due to the fact that a Couette stripe is approximately twice as wide as a Poiseuille stripe). Streamwise velocity fluctuations in the respective planes are shown in \figref{isolated-stripe-cuts} panels (a) to (e) for \ppf{} and in (f) to (j) for \pcf{}. In the latter case the largest fluctuation amplitudes are found for $\Delta=0$ whereas for \ppf{} fluctuation maxima are closer to $\Delta=5$. We will subsequently focus on these respective $z'$ locations, and in the wall-normal direction we select the crossing points between the laminar and turbulent profile, which in \pcf{} flow corresponds to $y=0$ and in \ppf{} to $y=0.46$. The streak dynamics at these respective locations are then shown in the spacetime plots in \figref{spacetime}, corresponding to one trajectory for each of \ppf{} and \pcf{}. We provide plots from one more trajectory for each system in the appendix (\figref{spacetime2}).
 
For \pcf{} we find that streaks are often wavy, implying that their spanwise position or alignment changes in time. New streaks arise at varying locations throughout the stripe, and these new streaks tend to push their neighbours apart causing them to alter their orientation. In \ppf{} on the other hand world lines remain straight, indicating that streaks keep their spanwise location (up to streak width) and preserve their alignment (see figures \ref{fig/spacetime}(a), \ref{fig/spacetime2}(a), and Supplementary Movie 2). Moreover, new streaks are only created locally at the downstream tip and the rate of streak creation is constant as can be seen from the constant slope of the downstream tip's trajectory in figures \ref{fig/spacetime}(a) and \ref{fig/spacetime2}(a). This local streak creation has been previously connected to an inflectional instability \citep{xiao2019growth}. The deterministic nature of stripe internal growth in \ppf{} was also pointed out in the experiments of \cite{mukund2021aging}. Furthermore, while in \pcf{} flow both ends of the stripe show similar irregular (/stochastic) behaviour, in \ppf{} the downstream and upstream tips behave differently. Whereas the downstream tip is perfectly regular (/deterministic), the upstream tip is only diffusive (no creation of streaks) and streaks are lost at irregular intervals, with multiple streaks sometimes disappearing at the same time, as can be seen in figures \ref{fig/spacetime}(a,b) and \ref{fig/spacetime2}(a,b). The above observations thus suggest that the proliferation process of individual stripes is stochastic in \pcf{} and deterministic in \ppf{}.
 
In \pcf{}, \cite{couliou2016spreading,couliou2015thesis} investigated the streak creation in spots in comparatively moderate aspect ratios (i.e.\ too small to encompass stripes). They distinguished between streak creation in the core of the spot (`inside streaks') and at the spot's boundary (`outside streaks'). For the streak creation in stripes investigated here, the streaks appear to be created more or less uniformly across the stripe. In the artificially restricted domain of \cite{duguet2011stochastic}, advection by the secondary flow is suppressed and nucleation of new streaks as well as decay events only happen at the tips.
 
In order to quantify the different nature of the stripe internal growth mechanism in \pcf{} and \ppf{}, a statistical description of the streak nucleation process is needed. To this end, we reproduce the spacetime diagrams of \figref{spacetime}(a,c) with a binary colour scheme in \figref{spacetime}(b,d) (and analogously for the second set of stripes shown in \figref{spacetime2}). Here, anywhere with positive/negative streamwise velocity fluctuation is shown in red/blue, and we mark the nucleation events with white dots. A newly created streak is marked only if it survives for at least 10 advective time units (as in \cite{couliou2016spreading,couliou2015thesis}). In \pcf{} flow nucleation events may happen `randomly' at any point in space and time, while in \ppf{} flow new streaks are generated only at the downstream tip.
 
It should be pointed out that in \ppf{} it is difficult, from the spacetime diagrams, to completely rule out streak generation inside the stripe. To ascertain this, we have drawn two inclined lines on top of figures \ref{fig/spacetime}(a) and \ref{fig/spacetime2}(a) bounding the downstream and upstream fronts (excluding the diffusive tail in the latter case). Since these two lines are parallel to each other, there appears to be no net growth in this core region. This can also be seen in Supplementary Movie 3 showing streamwise velocity fluctuations in \ppf{} in the frame of reference co-moving with the downstream tip. Here, it is clear that streaks are created at the downstream tip, with no apparent creation of new streaks elsewhere, and travel upstream along the stripe. Given that streaks remain by and large parallel as inferred from figures \ref{fig/spacetime}(a) and \ref{fig/spacetime2}(a), we can conclude that indeed there is no creation of new streaks inside the stripe's interior.
 
% Determinstic vs Stochastic
\begin{figure}
    \begin{overpic}[width=0.99\linewidth]{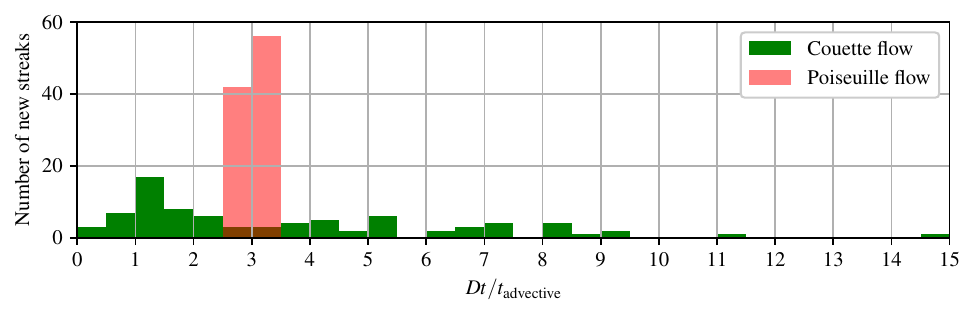}
    \put (-3, 30)  {(a)}
    \end{overpic}
    \begin{overpic}[width=0.99\linewidth]{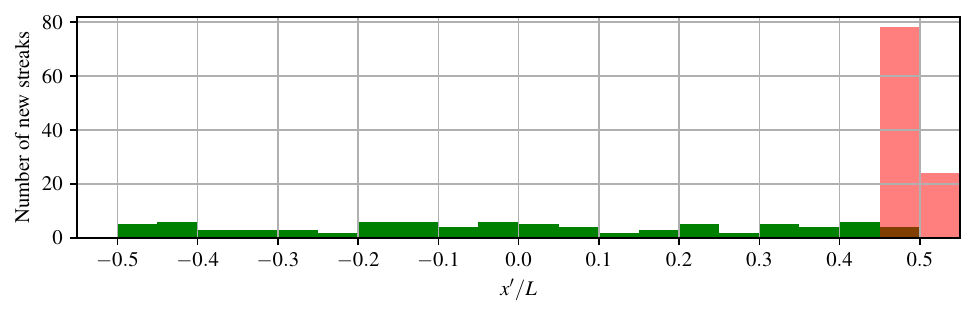}
    \put (-3, 30)  {(b)}
    \end{overpic}
    \caption{(a) Histogram of time between streak creation events in \ppf{} and \pcf{}. Bins are 0.5 advective time units each. Time of events are rounded to 1 simulation time, which equals $1/3$ advective time units for \ppf{} and $1/2$ advective time units for \pcf{}. (b) Histogram of the stripe-parallel location of streak creation events relative to stripe length ($L$, see \eqref{inst-length} for its definition). Bins are $0.05 x'/L$ each. For \ppf{}, the mean drift of the downstream tip [see dashed lines in figures \ref{fig/spacetime}(a) and \ref{fig/spacetime2}(a)] is subtracted, and its initial location is set to $x'/L=0.5$.}
    \label{fig/distribution}
\end{figure}
 
We record the time intervals $D t$ between nucleation events, marked in figures \ref{fig/spacetime}(b,d) and \ref{fig/spacetime2}(b,d), and we compute the distribution of such inter streak-creation times, with the result shown in \figref{distribution}(a). In \ppf{} the distribution has a distinct peak close to $D t / t_\mathrm{advective} \approx 3$ advective time units, confirming the regular, deterministic nature of the process. In \pcf{} on the other hand the distribution is broadband, thus showing the fundamentally different nature of the streak production underlying the internal stripe sustenance and proliferation process in \ppf{} and \pcf{}. This qualitative difference is equally apparent when we plot the distribution of the location of streak creations along the stripe [\figref{distribution}(b)]. In \pcf{} the number of streaks created is approximately constant along the entire stripe, whereas for \ppf{} streaks are exclusively created at the downstream tip. Note that to obtain this data two stripes were investigated for each geometry, counting in total 84 new streaks in \pcf{} and 102 in \ppf{}.

\subsection{Internal structure and transport}\label{sec/internals}
\label{sec/transport}
\begin{figure}
    \begin{overpic}[width=0.99\linewidth]{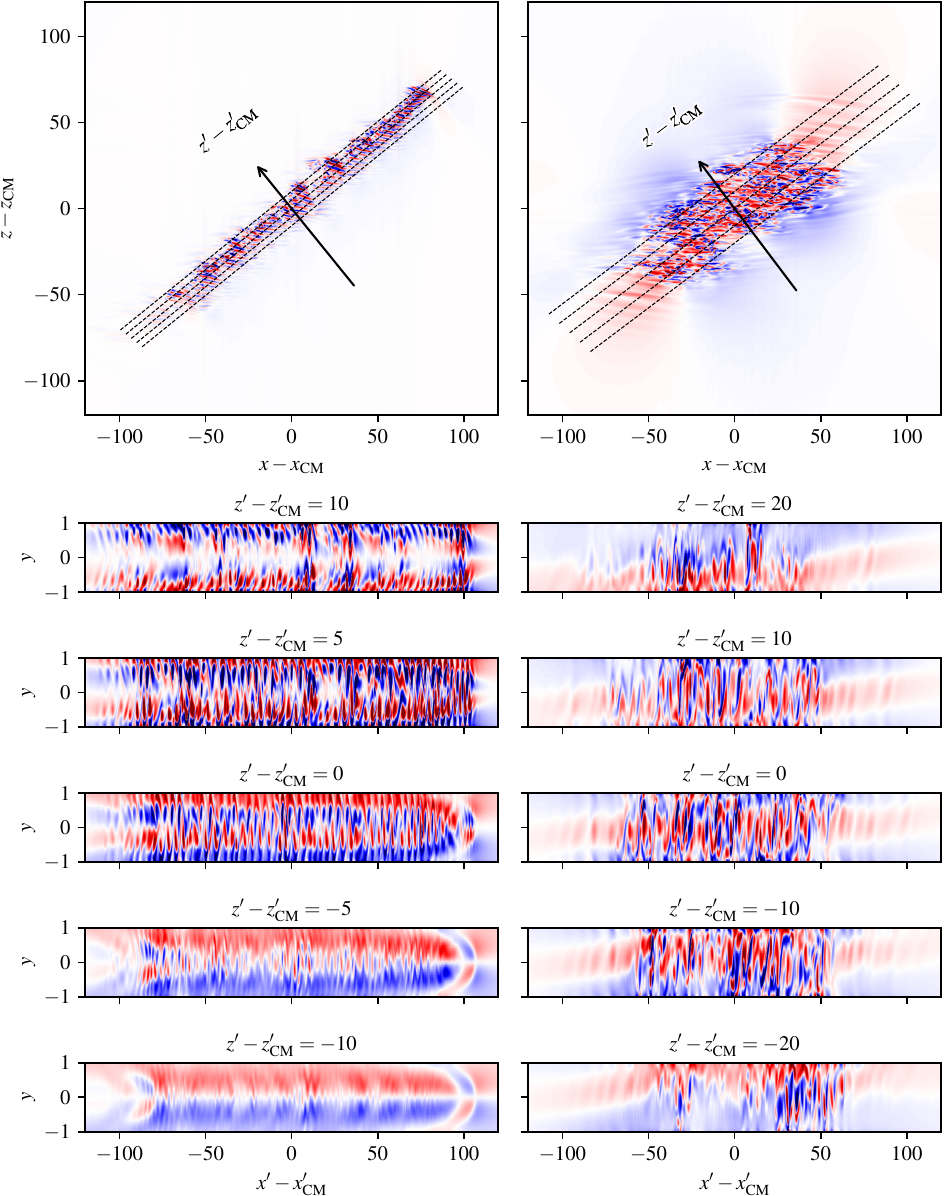}
        \put (0, 101)  {(a)}
        \put (0, 57)  {(b)}
        \put (0, 46)  {(c)}
        \put (0, 35)  {(d)}
        \put (0, 24)  {(e)}
        \put (0, 12)   {(f)}

        \put (42, 101)  {(g)}
        \put (42, 57)  {(h)}
        \put (42, 46)  {(i)}
        \put (42, 35)  {(j)}
        \put (42, 24)  {(k)}
        \put (42, 12)   {(l)}
    \end{overpic}
\caption{Streamwise vorticity $\omega_x$ in (a-f) \ppf{} and (g-l) \pcf{}. Top row is at the midplane, dashed lines indicate the locations of the stripe-perpendicular cuts in the rows below. The centres of mass ($x'_\mathrm{CM}, z'_\mathrm{CM}$) were marked in \figref{isolated-stripe}. Colours are capped at \(\omega_x = \pm 0.5\) for \ppf{} and \(\omega_x =\pm 1\) for \pcf{} for visibility.}
\label{fig/inclined-cuts}
\end{figure}
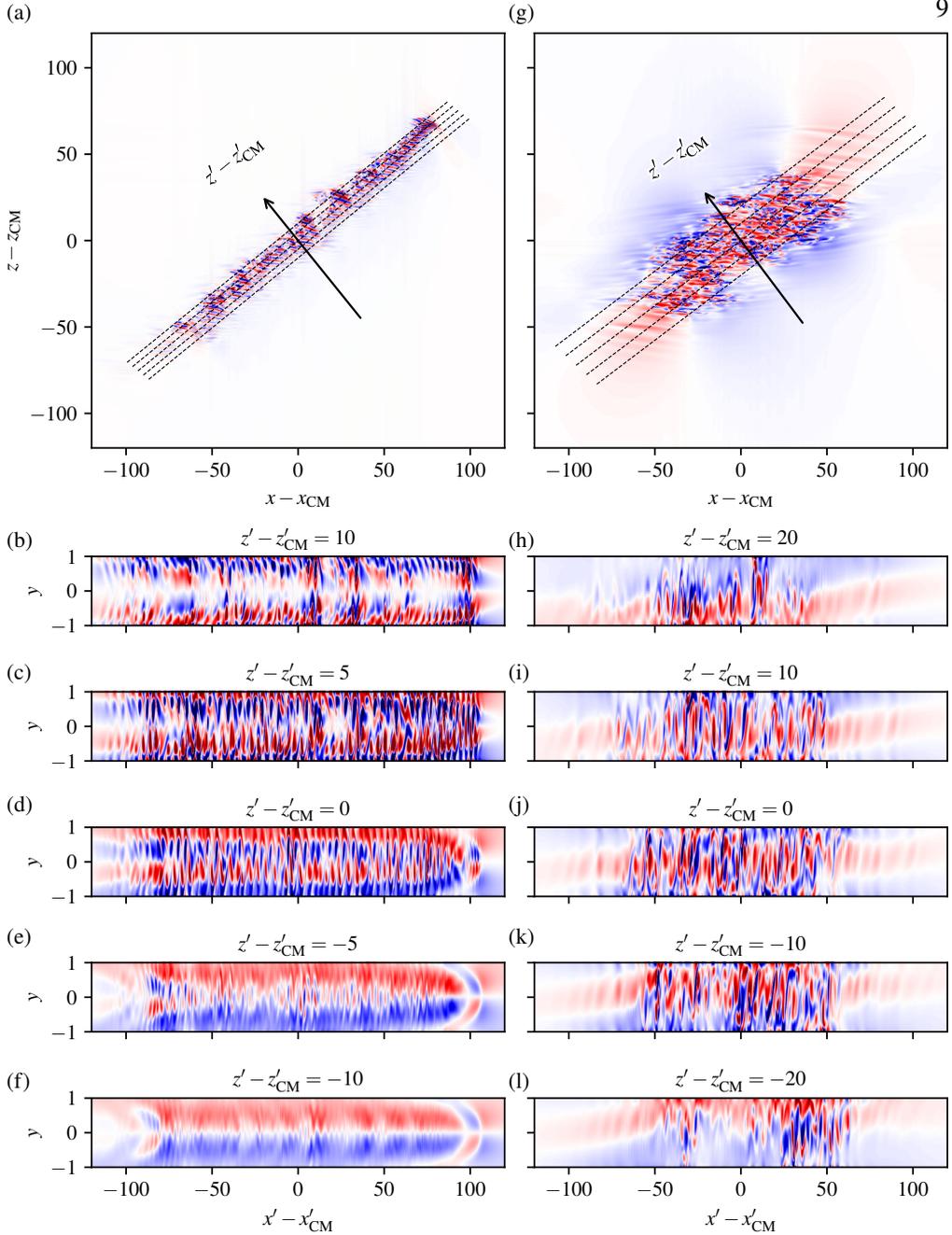
 
Given that the streak generation is markedly different between the two types of stripes we will in the following have a closer look at the stripes' interior with a specific focus on vorticity. In order to compare the internal structure of Couette and Poiseuille stripes, we first take cuts along stripes at the stripe-perpendicular locations indicated by the dashed lines in the top row of \figref{inclined-cuts} and visualize the streamwise vorticity field $\omega_x(x',y)$. Unexpectedly for \ppf{} (left column) the vortex patterns observed are close to regular, in particular for $z' - z'_{\mathrm{CM}} \le 0$ (panels d-f). Although fluctuations are non-zero in this downstream area, the flow pattern looks far from what one would ordinarily consider as turbulent. Also, this feature is very different from the \pcf{} case (\figref{inclined-cuts}, right column), where the pattern is much less regular throughout the stripe.
 
Looking separately at the two terms that compose streamwise vorticity, i.e.\ $\partial u_y/\partial z$ and $\partial u_z/\partial y$ (not shown), we find that the regular pattern of vorticity at the downstream side of the Poiseuille stripe is due only to the latter term, i.e.\ the spanwise shear. \cite{xiao2019growth} attributed the mechanism of streak generation at the downstream tip of a localized stripe in \ppf{} to a spanwise inflectional instability of the local mean flow. The latter was computed by temporally and spatially averaging the velocity field obtained from DNS at $\Rey^P=750$. Here, we observe a flow pattern similar to that of the most unstable perturbation found in their linear stability analysis (see their figure 8). Our observation hence suggests that the same spanwise inflectional instability mechanism remains central to stripes at Reynolds numbers closer to the critical point.
 
 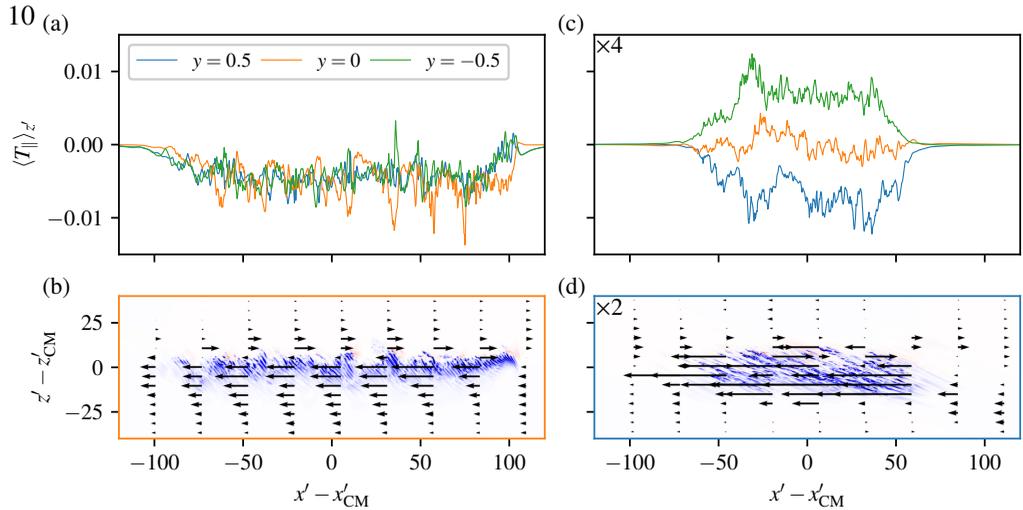
\begin{figure}
     \begin{overpic}[width=0.99\linewidth]{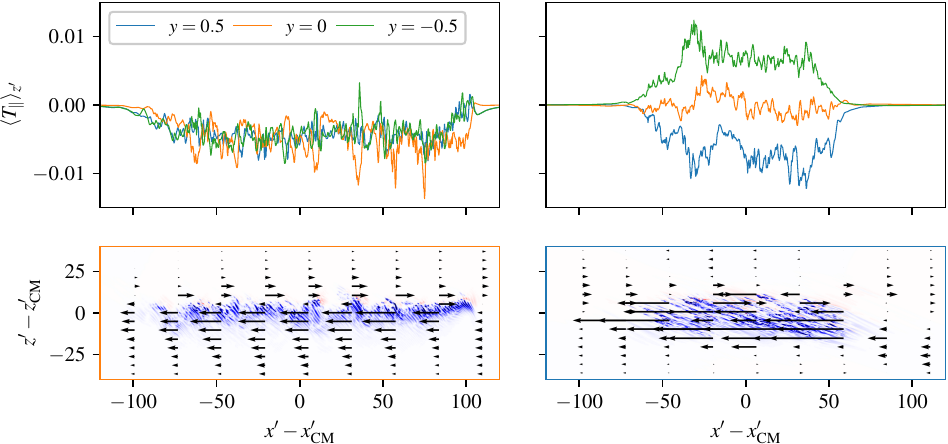}
         \put (3, 47)  {(a)}
         \put (54, 47)  {(c)}
         \put (3, 21)  {(b)}
         \put (54, 21)  {(d)}
         \put (57.75, 44.75)  {$\times 4$}
         \put (57.75, 19)  {$\times 2$}
     \end{overpic}
   \caption{Stripe-parallel transport of streamwise vorticity ($T_\parallel=|\omega_x|\,u_{x'} $) in (a,b) \ppf{} and (c,d) \pcf{}. Top row shows the average of $T_\parallel$ in the stripe-perpendicular direction $z'$ near the stripes ($z' - z'_\mathrm{CM} \in [-20,20]$ for \ppf{} and $z' - z'_\mathrm{CM} \in [-40,40]$ for \pcf{}) at different wall-normal planes. Bottom row shows $T_\parallel$ (colours) and the stripe-parallel velocity fluctuations $u_{x'}$ (arrows). $T_\parallel$ is shown at (b) \(y=0\) for \ppf{} and (d) \(y=0.5\) for \pcf{} (bounding box of each plot follows the same colour-coding as in the top row). Displayed arrows are proportional, with the same scale in (b,d), to the stripe-parallel velocity fluctuations $u_{x'}$ averaged in the whole wall-normal extent for \ppf{} (b), and in the upper-half wall-normal extent for \pcf{} (d). Colours are capped at \(T_\parallel = \pm 0.1\) for \ppf{} and \(T_\parallel = \pm 0.4\) for \pcf{} for visibility. Ticks on the vertical axis of (c) have values four times the values of the corresponding ticks of (a). Similarly, vertical ticks of (d) have double the values of the ticks in (b).}
   \label{fig/transport-stripeparallel}
 \end{figure}
 
We will next consider the stripe internal vortex dynamics which at least in part will be linked to the secondary flow. Here, we refer to the deviation of the velocity field from laminar flow in the streamwise-spanwise plane as the secondary flow. This is a three-dimensional field and its wall-normal average was shown by the arrows in \figref{isolated-stripe}. We compare the transport of streamwise vorticity along the stripe in \pcf{} and \ppf{} by computing $T_\parallel=|\omega_x|\,u_{x'} $, where $u_{x'}$ is the stripe-parallel velocity fluctuations. The stripe-parallel transport averaged over the stripe-perpendicular direction, i.e.\ $\langle T_\parallel \rangle_{z'}$, is shown at different wall-normal planes in \figref{transport-stripeparallel}(a) and (c) for \ppf{} and \pcf{}, respectively. In \ppf{} vortices travel from their creation point at the downstream tip to the upstream tip, where they eventually die out. The resulting stripe-parallel transport is thus negative at all wall-normal locations. This transport direction is also evident in \figref{transport-stripeparallel}(b) in the arrows that show the average of the stripe-parallel velocity fluctuations $u_{x'}$ in the whole wall-normal extent. It is noteworthy that this transport direction coincides with the secondary flow at the stripe's downstream side. As can be seen in \figref{transport-stripeparallel}(b), the stripe-parallel secondary flow on the stripe's downstream side is stronger than that on the upstream side. 
 
In Couette flow vortices travel in opposite directions in the upper and lower planes, and at the stripe's midplane ($y=0$), the stripe-parallel transport fluctuates around zero, see \figref{transport-stripeparallel}(c). If we separately consider the flow in the upper and lower domain half ($y>0$ and $y<0$), then for each subdomain, up- and downstream are defined by the nearest wall motion and the resulting mean advection. In this case the direction of vortex motion in each subdomain again follows the secondary flow at the stripe's downstream interfaces as can be seen from \figref{transport-stripeparallel}(d). Here the stripe-parallel secondary flow is averaged in the upper-half wall-normal extent of the domain. (The corresponding plot in the lower half qualitatively matches the reflection of \figref{transport-stripeparallel}(d) around $x'- x'_\mathrm{CM}=0$ and $z'- z'_\mathrm{CM}=0$ followed by the negation of $u_{x'}$ and $u_{z'}$.) Also in this case the secondary flow is stronger at the downstream interface and here has the same direction as in the stripe's interior while at the upstream interface the secondary flow has the opposite orientation and is much weaker. The direction of the streak advection may be considered counterintuitive given that this motion is partially opposed to the mean flow advection along the stripe in each subdomain.
 
Comparing the magnitudes of vorticity transport between \figref{transport-stripeparallel}(a) and (b), it is apparent that this quantity is approximately four times stronger in \pcf{} than it is in \ppf{}. Returning to the stripe visualizations of \figref{isolated-stripe}, the streamwise fluctuations are also stronger in \pcf{}, in this case by a factor of two. Correspondingly disturbance kinetic energy levels ($E = \langle u^2+v^2+w^2 \rangle / 2$) differ by a factor of four between the two flows (as will also be seen later in \figref{fronts-avg-v2-E}). Our observation is in line with recent findings \citep{andreolli2021global} of larger energy as well as stronger secondary flow in fully turbulent \pcf{} as compared to turbulent \ppf{}. It should be noted, however, that a different scaling of the velocity and lengths, based on the average shear, has been proposed by several other studies \citep{barkley2007mean, tsukahara2010turbulence} in order to compare Poiseuille and Couette flows. By viewing Poiseuille flow as a superimposition of two shear layers \citep{tuckerman2020patterns}, the characteristic velocity used for the non-dimensionalization turns out to be twice that of Couette flow. Therefore, with such a scaling, streamwise fluctuations and disturbance kinetic energy levels in our simulations would have comparable magnitudes between the two flows.
 
% fig E and v^2
\begin{figure}
    \begin{overpic}[width=0.99\linewidth]{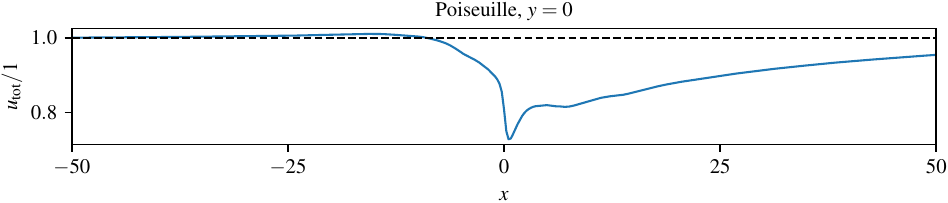}
        \put (0,20) {(a)}
    \end{overpic}\\
    \vspace{2mm}
    \begin{overpic}[width=0.99\linewidth]{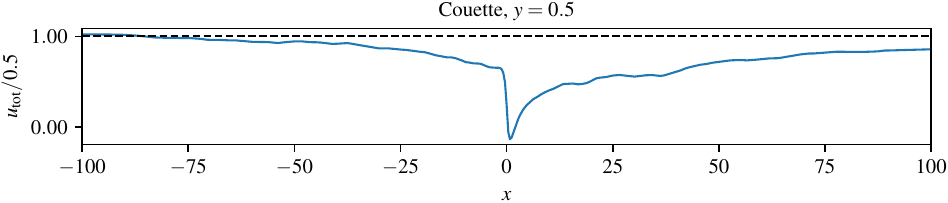}
        \put (0,20) {(b)}
    \end{overpic}\\
    \vspace{2mm}
    \begin{overpic}[width=0.99\linewidth]{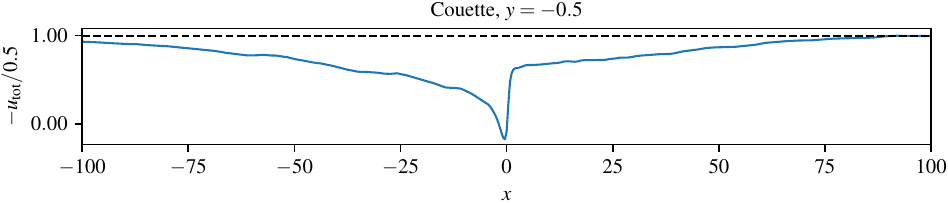}
        \put (0,20) {(c)}
    \end{overpic}
    \caption{Time averages of total streamwise velocity (normalized by the laminar values) at (a) midplane in \ppf{} and (b,c) upper/lower-half planes in \pcf{}, at the spanwise centres of mass. Each time series was averaged individually, shifting the location of the maximum of \(|dE/dx|\) at the given wall-normal plane to \(x=0\) prior to the time averaging. Here, and in figures \ref{fig/fronts-avg-v2-E}-\ref{fig/fronts-avg-profiles}, time averages are performed over two different trajectories for each geometry, and each trajectory is averaged for 300 and 600 advective time units for \pcf{} and \ppf{}, respectively. We have verified that these choices provided sufficient statistics to produce robust results.}
    \label{fig/fronts-avg-u}
\end{figure}
 
% fig tot streamwise velocity
\begin{figure}
    \begin{overpic}[width=0.99\linewidth]{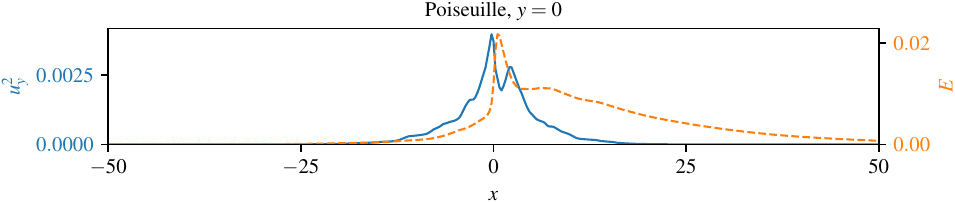}
        \put (0,19) {(a)}
    \end{overpic}\\
    \vspace{2mm}
    \begin{overpic}[width=0.99\linewidth]{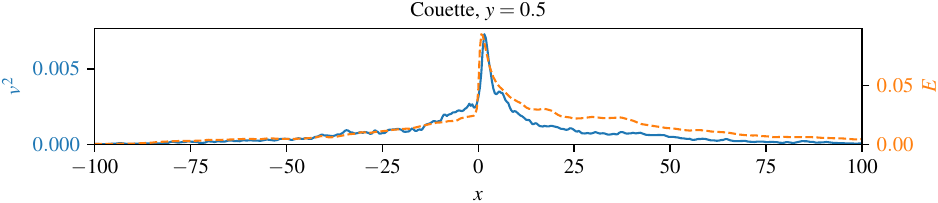}
        \put (0,19) {(b)}
    \end{overpic}\\
    \vspace{2mm}
    \begin{overpic}[width=0.99\linewidth]{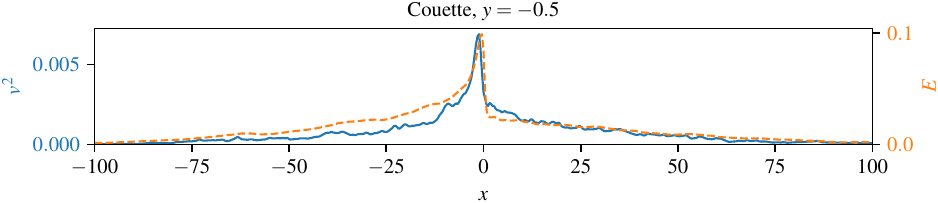}
        \put (0,19) {(c)}
    \end{overpic}
    \caption{Time averages of squared wall-normal velocity (solid lines) and disturbance kinetic energy (dashed lines) at (a) midplane in \ppf{} and (b,c) upper/lower-half planes in \pcf{}, at the spanwise centres of mass. Each time series was averaged individually, shifting the location of the maximum of \(|dE/dx|\) at the given wall-normal plane to \(x=0\) prior to the time averaging.}
    \label{fig/fronts-avg-v2-E}
\end{figure}
 
The internal transport processes in Couette flow suggest a separation between the upper and lower half domains, which of course is also implicit given the up/down anti-symmetry of the boundary conditions. We will use such an independent consideration of the two half domains in the following in order to better understand the secondary flow along the fronts in the upper and respectively lower half domain. Our analysis of secondary flows will be less rigorous than the full explanation provided by \cite{duguet2013oblique}, yet it provides an explanation for the presence of a stronger flow component along the stripes' downstream turbulent-laminar interface (and within the overhang region) when compared to the upstream interface. Before we more closely look at these interfaces in \pcf{} and \ppf{}, we briefly summarize recent studies of front dynamics in pipe flow.

\subsection{Laminar-turbulent interfaces}\label{sec/interfaces}
As discussed in the introduction, puffs in pipe flow rely on energy input from the upstream parabolic flow, whereas the downstream plug flow has to recover towards a parabolic shape before a second puff can be sustained. Accordingly, puffs are composed of two different types of interfaces (/fronts). At the upstream laminar-turbulent interface the velocity profile sharply changes from an energetic parabolic profile to a plug flow. The downstream interface on the other hand is characterized by a slow profile adjustment and its recovery back to the parabolic shape. This process relies on the action of viscosity, starting from the wall, and typically the profile only regains a sufficiently parabolic shape at a downstream distance of approximately 25 diameters \citep{samanta2011experimental}. Puffs can hence not merge to form larger entities, they adhere to a minimum spacing of $\sim25$ diameters.
 
As pointed out by \cite{barkley2011simplifying,barkley2016theoretical}, pipe flow can be interpreted as an excitable bistable medium, where the laminar flow corresponds to the excitable rest state, turbulence to the excited state and the fast and slow timescales to the excitation of turbulence and the viscous profile recovery back to the parabolic flow, respectively.
 
In planar Couette and Poiseuille flow, stripe inclination and the corresponding secondary motions complicate matters. It is nevertheless reasonable to assume that in analogy to pipe flow localization is connected to a viscous recovery process at the respective downstream laminar turbulent interfaces. The spacing between Couette stripes is in fact comparable to the minimum spacing between puffs \citep{samanta2011experimental}.
 
For Poiseuille stripes, the corresponding weak and strong fronts can be readily identified by the customary centreline velocity plot. \Figref{fronts-avg-u}(a) shows the characteristic sharp drop at the upstream interface, marking it as a strong front, whereas at the downstream side the profile only gradually recovers, marking it as a weak front. For the time averaging, since stripes (like puffs) do not have a perfectly constant speed, we chose the x-location with the steepest increase in disturbance kinetic energy as reference point. The corresponding time averaged energy is shown as a function of the streamwise coordinate in \figref{fronts-avg-v2-E}.
 
% Profiles
\begin{figure}
    \begin{overpic}[width=0.99\linewidth]{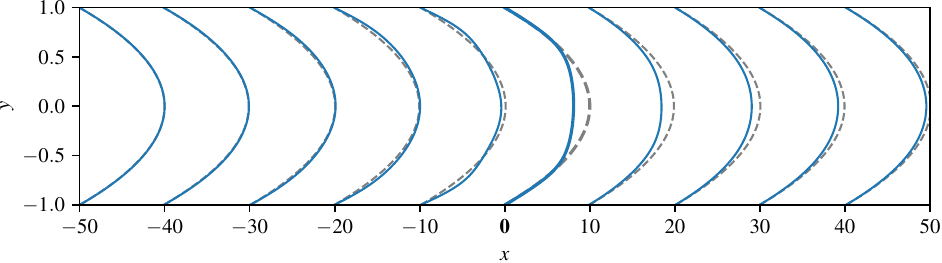}
        \put (0,26.5) {(a)}
    \end{overpic}\\
    \vspace{2mm}
    \begin{overpic}[width=0.99\linewidth]{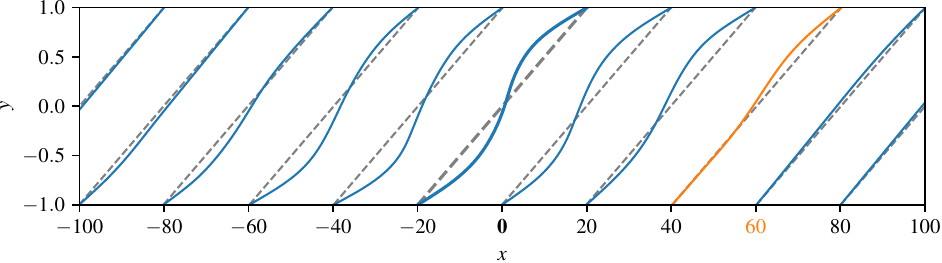}
        \put (0,28) {(b)}
    \end{overpic}
\caption{Time averages of velocity profiles (solid lines) of (a) \ppf{} and (b) \pcf{} at the spanwise centres of mass. Each the time series was averaged individually, shifting the streamwise location of the centre of mass to \(x=0\) prior to the time averaging. Ticks on the $x$ axes are at the locations of zero velocities of the laminar profiles (dashed), we highlighted the profiles at $x=0$ with bolder linewidths for clarity. See the text for a discussion where the profile at $x=60$ in (b), highlighted in orange, is used.}
    \label{fig/fronts-avg-profiles}
\end{figure}

In \pcf{}, as discussed in the previous section, we will consider the upper- and lower-half domains separately. In either half indeed a sharp velocity drop is observed at the stripe's `upstream' interface and a slower profile adjustment back to laminar downstream. To illustrate this we select a wall-normal location of $y=0.5$. Like for \ppf{}, we pick the location of the steepest rise in disturbance kinetic energy as the reference location for time averaging. The same procedure is repeated for $y=-0.5$. It should be noted that the location of the steepest rise in $E$ in the lower half is not necessarily correlated to the respective location in the upper half and both move with respect to each other. Hence, the averaging processes in the half domains are carried out independently. The total streamwise velocity and the disturbance kinetic energy are shown in \figref{fronts-avg-u} and \figref{fronts-avg-v2-E}. Couette stripes have hence two weak diffusive interfaces extending in opposite directions in the upper and lower halves respectively (see Figure 5 in \cite{barkley2011modeling} for an illustration). The wall-normal component of disturbance kinetic energy, which is a measure of turbulence intensity, in models commonly denoted as `$q$' \citep{barkley2011simplifying,barkley2016theoretical}, and which corresponds to the fast timescale of fluctuations, is separately plotted (blue curves) in \figref{fronts-avg-v2-E}. In \ppf{}, as in pipes, this quantity drops off fast while the centreline velocity only recovers more slowly. This perfectly reflects the separation of timescales and the flow's refractory nature, a central aspect in the context of excitable media. Also, the adjustment of the average velocity profile across the stripe \figref{fronts-avg-profiles}(a) qualitatively agrees with profiles across puffs in pipe flow. For Couette stripes, on the other hand, the energy in the wall-normal velocity [\figref{fronts-avg-v2-E}(b,c)] drops off much more slowly than for Poiseuille stripes [\figref{fronts-avg-v2-E}(a)]. The energy adjustment appears to occur on a timescale comparable to the recovery of velocity profile and the separation between timescales is not so clear. \cite{wang2022stochastic} attributed the different interface structures between pressure-driven and shear-driven flows to the different forcing mechanisms (energy is supplied by mean flow from the upstream direction in the former case, while in the latter energy is uniformly transported from the boundaries to the fluid inside through shear). Because of these differences, they argued that no energy depletion zone occurs in Couette flow. This is different however from what we observe, as we do find sharp and diffusive interfaces in both geometries, once the upper and lower halves are considered separately in Couette flow.

The circumstance that Couette stripes have two weak fronts and Poiseuille stripes only have one is also reflected by the differences in stripe width. As can be seen in \figref{isolated-stripe}, Couette stripes are approximately twice as wide as their Poiseuille counterparts.

Finally, we look at the local velocity profiles in \pcf{} and their development across the stripe. Again we will focus on just half the domain, in this case on the upper half of \figref{fronts-avg-profiles}. Sufficiently far upstream (left) as well as downstream (right) of the stripe the profiles assume their linear laminar shape. In the stripe's interior the profiles are curved and total streamwise velocity and its mean (in the upper-half plane) are considerably lower than that of the laminar flow. Let's now pick the upper-half velocity profile at $x=60$ in the right-hand side overhang region in \figref{fronts-avg-profiles}(b) (orange line). Here the streamwise flux is still considerably lower than in the laminar case, but at the same time the flow accelerates in the $x$ direction (i.e.\ the profile approaches the laminar one as we move in the positive $x$ direction). Because of continuity this mismatch in the streamwise upper-half velocity profiles must be compensated by a secondary flow which could be an inflow either from the lower half domain or from the sides (i.e.\ spanwise). The velocity profile at the same location $x=60$ in the lower half is however close to laminar, only leaving a spanwise flow in the upper half domain to account for the necessary compensation. At the upper-half upstream interface $x=-40$ the profile adjustment only partially occurs via the stripe parallel velocity component, at this location there is actually momentum transport between the upper and lower domain halves. Hence, the secondary flow on the stripes upstream interface is weaker than that at the downstream interface.
 
While we have discussed the internal streak creation of stripes, which is essential for their sustenance and elongation, their growth in the stripe-perpendicular dimension ($z'$) at low Reynolds numbers tends to occur in the form of stripe splitting. The name here is chosen in analogy to puff splitting in pipes, yet as we will see, depending on geometry, the processes can be quite different. In pipe flow as mentioned above vortices have to escape from the parent stripe and breach the downstream `refractory' zone. This escape naturally can only occur in the streamwise direction. Stripes on the other hand can separate in different ways.
 
\subsection{Splitting mechanism}\label{sec/splitting}
\begin{figure}
    \begin{overpic}[width=0.248\linewidth]{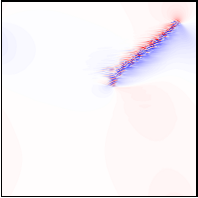}
        \put (-15,100) {(a)}
    \end{overpic}
    \begin{overpic}[width=0.248\linewidth]{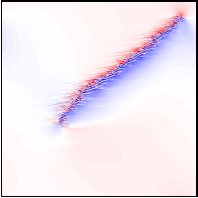}
        \put (85,105) {\large{Time $\rightarrow$}}
    \end{overpic}
    \begin{overpic}[width=0.248\linewidth]{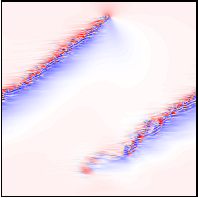}
    \end{overpic}
    \begin{overpic}[width=0.248\linewidth]{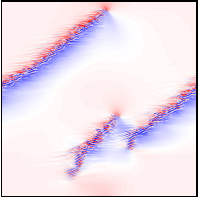}
    \end{overpic}\\
    \begin{overpic}[width=0.248\linewidth]{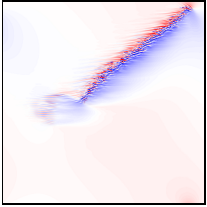}
        \put (-15,100) {(b)}
    \end{overpic}
    \begin{overpic}[width=0.248\linewidth]{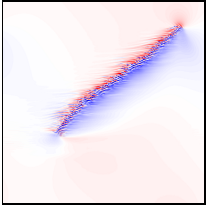}
    \end{overpic}
    \begin{overpic}[width=0.248\linewidth]{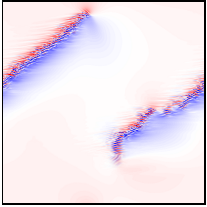}
    \end{overpic}
    \begin{overpic}[width=0.248\linewidth]{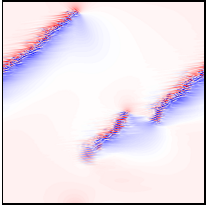}
    \end{overpic}\\
    \begin{overpic}[width=0.248\linewidth]{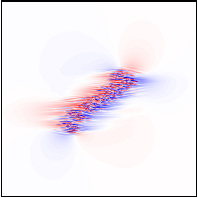}
        \put (-15,95) {(c)}
        \put (-25,100) {\rotatebox[origin=c]{90}{\large{$z \rightarrow$}}}
    \end{overpic}
    \begin{overpic}[width=0.248\linewidth]{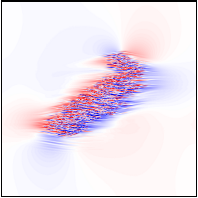}
    \end{overpic}
    \begin{overpic}[width=0.248\linewidth]{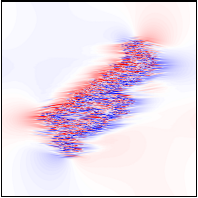}
    \end{overpic}
    \begin{overpic}[width=0.248\linewidth]{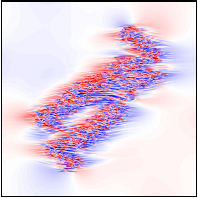}
    \end{overpic}\\
    \begin{overpic}[width=0.248\linewidth]{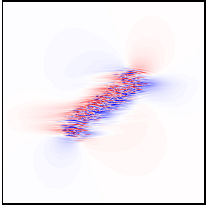}
        \put (-15,100) {(d)}
    \end{overpic}
    \begin{overpic}[width=0.248\linewidth]{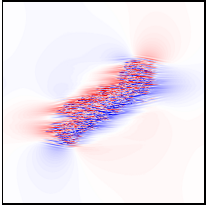}
        \put (95,-9) {\large{$x \rightarrow$}}
    \end{overpic}
    \begin{overpic}[width=0.248\linewidth]{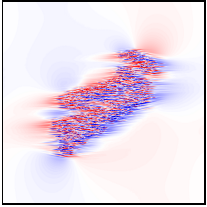}
    \end{overpic}
    \begin{overpic}[width=0.248\linewidth]{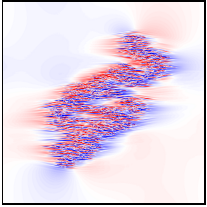}
    \end{overpic}\\
    \caption{Stripe evolution and proliferation in (a,b) \ppf{} at $\Rey^P=750$ and (c,d) \pcf{} at $\Rey^C=350$. Streamwise velocity fluctuations $u_x$ are visualized in the $x$-$z$ planes at $y=0.46$ in \ppf{} and $y=0$ in \pcf{}, respectively. See figures \ref{fig/splitting-more-channel} and \ref{fig/splitting-more-couette} for two more cases from \ppf{} and \pcf{}, respectively.}
    \label{fig/splitting}
\end{figure}
 
The purpose of this section is to highlight the qualitative differences observed for stripe splittings between \ppf{} and \pcf{}. Splitting here refers to the creation of new (daughter) stripes from an existing (parent) stripe. We focus on the splitting mechanism that is encountered in the vicinity of the critical point where stripe turbulence first becomes sustained in the respective geometry. Other ways of stripe creation may become relevant at higher \Rey{} \citep{manneville2012growth,manneville2020transitional}. As we will see, the splitting mechanism in \ppf{} differs fundamentally from that in \pcf{}, in that in the former case the stripe splits perpendicular to its length (the parent stripe becomes shorter), whereas in the latter case the splitting occurs along the stripe (the parent stripe becomes slimmer). Starting with \ppf{} we show two examples of a typical splitting event ($\Rey^P=750$) in \figref{splitting}(a,b) (see also two additional cases in \figref{splitting-more-channel}). The parent stripe grows until it eventually sheds its (upstream) tail. The shed part is located parallel and upstream of the parent stripe and grows into a daughter stripe. This process hence strongly depends on the elongation of the parent stripe. It is noteworthy that this splitting process is not memoryless, because the splitting probability turns out to be proportional to the length of the parent stripe and is hence not constant in time, as has been pointed out recently \citep{mukund2021aging}.
 
 \begin{figure}
     \begin{overpic}[width=0.99\linewidth]{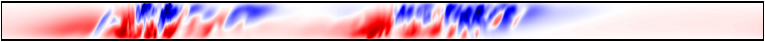}
     \end{overpic}\\
     \begin{overpic}[width=0.99\linewidth]{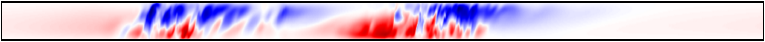}
     \end{overpic}\\
     \begin{overpic}[width=0.99\linewidth]{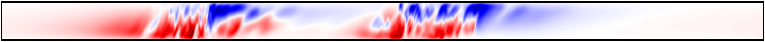}
     \end{overpic}\\
     \begin{overpic}[width=0.99\linewidth]{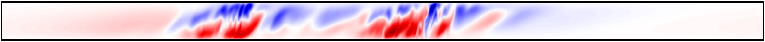}
     \end{overpic}\\
     \begin{overpic}[width=0.99\linewidth]{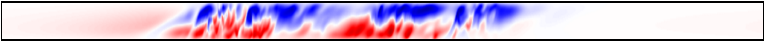}
     \end{overpic}\\
     \begin{overpic}[width=0.99\linewidth]{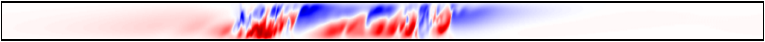}
     \end{overpic}\\
     \begin{overpic}[width=0.99\linewidth]{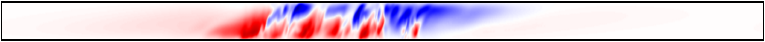}
         \put (-4,10) {\rotatebox[origin=c]{90}{\large{Time $\rightarrow$}}}
         \put (102,15) {\rotatebox[origin=c]{270}{\large{$\leftarrow y$}}}
     \end{overpic}\\
     \begin{overpic}[width=0.99\linewidth]{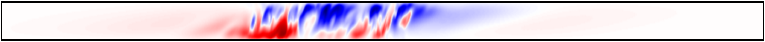}
     \end{overpic}\\
     \begin{overpic}[width=0.99\linewidth]{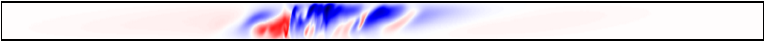}
     \end{overpic}\\
     \begin{overpic}[width=0.99\linewidth]{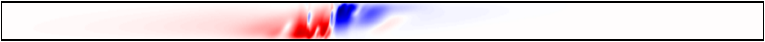}
         \put (43,-2) {\large{$x \rightarrow$}}
     \end{overpic}
    \caption{Splitting process in \pcf{} at $\Rey^C=350$ for the case shown in \figref{splitting}(c). Shown are the streamwise velocity fluctuations $u_x$ on the $x$-$y$ plane at the spanwise centres of mass. The splitting process relies on the broadening of the stripe, which appears to be initiated at one side of the stripe (via the respective weak front).}
 \label{fig/splitting-to-the-right}
 \end{figure}

Two examples of stripe splitting in \pcf{} ($\Rey^C=350$) are shown in \figref{splitting}(c,d) (see also two additional cases in \figref{splitting-more-couette}). In this case the original stripe not only grows in length but, more importantly, also broadens as shown in \figref{splitting-to-the-right}. Broadening appears to be initiated at one side of the stripe (via the respective weak front) and once the stripe has approximately doubled in width a gap forms along the stripe, splitting it into two slimmer versions. In this case, and hence in contrast to \ppf{}, the growth in the stripe-perpendicular direction leads to the creation of a new stripe. This process overall appears similar to a recent proposition by \cite{frishman2022mechanism} for puff splitting in pipe flow. Here it has been suggested that puff splitting can be regarded as a precursor of the puff-slug transition, i.e.\ the uniform expansion of turbulence that is only stable at larger $\Rey$. Driven by fluctuations puffs that normally would not expand can briefly tap into the slug growth mechanism. Yet after a short expansion period the refractory nature re-emerges and the now-too-broad puff develops a gap in the middle and splits into two. It is noteworthy that the splitting process for short periodic stripes in Couette flow \citep{shi2013scale} is in principle the same as the one observed here for fully localized Couette stripes. In both instances the splitting occurs parallel to the stripe direction. Interestingly, splittings of short periodic Poiseuille stripes \citep{gome2020statistical} equally occur parallel to the stripe. Splitting perpendicular to the stripe length is hence only encountered for localized Poiseuille stripes, and we argue that it is intimately linked to the downstream tip instability and the associated continuous stripe elongation. In this case the streaks created at the downstream tip are transported to the upstream tip, where shedding events give rise to daughter stripes. Conversely, in Couette flow, as also stated above, splitting is not driven by stripe elongation but by broadening of the stripe followed by gap formation. In this case the stripe internal streak transport occurs also along the stripe but in opposite directions in the upper and lower halves of the domain. The broadening of the stripe appears to be driven by the weak front, i.e.\ it tends to start at the `downstream' laminar-turbulent interface in the respective half-domain.

\section{Conclusion}\label{sec/conclusions}

We have seen that Poiseuille and Couette stripes, despite being similar in appearance, substantially differ with respect to the internal growth and transport processes. One of the most peculiar aspects is the deterministic streak (/vortex) generation encountered for Poiseuille stripes and the close-to-regular internal stripe structure that results from it. At the low Reynolds numbers considered here Poiseuille stripes are entirely dominated by this regular growth at the downstream tip and the advection of streaks towards the upstream tip. As has been proposed in a recent study \citep{mukund2021aging} this mechanism appears to persist for $\Rey \gtrapprox 650$ and leads to the elongation and sustenance of individual stripes. This observation is in agreement with an earlier computational study \citep{Tao2018extended} where the authors observed individual stripes to become sustained at $\Rey=660$, but these authors argued at the same time that this mechanism only gives rise to sparse turbulence and that spreading in two dimensions only occurs at a second critical point where bands split ($\Rey \approx 1000$). In experiments \citep{mukund2021aging} where instead of a few, many thousand stripes have been investigated, splittings have been shown to already occur at $\Rey \approx 675$ and hence only slightly above the point where stripes first begin to expand. Moreover, as we discussed in \secref{splitting}, splitting of Poiseuille stripes relies on the streamwise extension and follows from a subsequent shedding of the upstream tail (see \figref{splitting}(a,b) and also \figref{splitting-more-channel} for two more cases). The probability of such sheddings to occur increases with stripe length \citep{mukund2021aging}. This connection between proliferation of individual stripes and the creation of new stripes can be regarded as an argument in favour of a single critical point where stripe turbulence in Poiseuille flow first becomes sustained.

As pointed out by \cite{shimizu2019bifurcations} this splitting process only gives rise to daughter stripes that have the same orientation as the parent stripe. Couette stripes on the other hand grow by stochastic streak creation that occurs throughout the bulk of the stripe and do not depend on the tip dynamics. Splitting then does not follow from stripe extension but rather from a broadening and gap formation along the stripe. While we limited our study to splitting, it should be mentioned that the broadening of Couette stripes can equally give rise to branches that grow perpendicular to the original stripe, which then may or may not break off, and hence stripes of both orientations can be created in the process. As we argued, in the case of Couette stripes it is useful to distinguish between the upper- and lower-half domains, where the weak and strong fronts can be readily identified, and the broadening and proliferation of the respective half of the stripe occurs in the direction of the mean advection in the respective domain. 

Past studies considered stripe splitting in tilted domain Couette \citep{shi2013scale} and Poiseuille \citep{gome2020statistical} simulations, in configurations where stripes connect via the shorter periodic direction and can develop neither upstream nor downstream tips. In these cases, splittings can only occur in the stripe perpendicular direction and hence look similar (see fig.\ 4d in \cite{shi2013scale} and fig.\ 10 in \cite{gome2020statistical}) to the large domain Couette splittings observed in the present study. It is noteworthy that in Poiseuille experiments and large aspect ratio simulations \citep{paranjape2019thesis}, broadening of stripes and branching events are found but only at considerably larger Reynolds numbers ($\Rey>1000$). It is reasonable to assume that once stripes broaden splittings in the stripe perpendicular direction (at the downstream interface) may also occur. We hence suspect that at higher Reynolds numbers two different splitting mechanisms may be at play in Poiseuille flow. For the sustenance of Poiseuille stripes the mechanism encountered at lower Reynolds numbers is the one that matters, which is the splitting mechanism we describe in the present study. As this mechanism relies on the streak creation at the downstream tip, it is unique to Poiseuille flow and requires stripes to be fully localized.

With respect to the nature of the transition in the two flows, here the key difference is again the regular, deterministic internal growth in Poiseuille stripes compared to the stochastic growth in Couette stripes. While for Couette flow the transition falls into the directed percolation universality class \citep{klotz2022phase}, stochasticity is a key feature of this transition type and the peculiarities of Poiseuille stripes may potentially lead to a transition of a different nature, as has been argued by \cite{shimizu2019bifurcations,mukund2021aging}. 

Finally, we would like to briefly mention recent studies in Couette--Poiseuille flow \citep{klotz2017couettepoiseuille,morimatsu2020laminartextendash,klotz2021experimental,liu2021decay,shuai2022structured}, a geometry that allows to link the two flows studied here. Future studies in this geometry could investigate the nature of stripes for increasing pressure driving, i.e.\ continuously approaching the Poiseuille limit. We would anticipate that at a critical driving, streak production may set in at the downstream tip. How the stripe's internal structure changes at this point and how in turn the splitting mechanism is affected by these internal changes would provide valuable insights into the differences between the two limiting cases.

\backsection[Supplementary data]
    {Supplementary movies are available at [URL will be inserted by the publisher.]}
\backsection[Acknowledgements] 
    {We would like to thank Ashley Willis for sharing his couette-channel flow solver and Chaitanya Paranjape and Mukund Vasudevan for sharing some of the plane-Poiseuille flow data. EM would like to thank Sébastien Gomé and Yohann Duguet for their useful comments 
    and suggestions on preliminary versions of the manuscript.}
\backsection[Funding]
    {E.\ Marensi acknowledges funding from the ISTplus fellowship programme.
    G.\ Yalnız and B.\ Hof acknowledge a grant from the Simons Foundation 
    (662960, BH).}
\backsection[Declaration of interests]{The authors report no conflict of interest.}
\backsection[Author ORCID]{
    E.\ Marensi, \url{https://orcid.org/0000-0001-7173-4923};
    G.\ Yalnız, \url{https://orcid.org/0000-0002-8490-9312}; 
    B.\ Hof, \url{https://orcid.org/0000-0003-2057-2754}}

\appendix

\setcounter{equation}{0}
\setcounter{figure}{0}
\setcounter{table}{0}

\renewcommand{\theequation}{A\arabic{equation}}
\renewcommand{\thefigure}{A\arabic{figure}}
\renewcommand{\thetable}{A\Roman{table}}

\section{}\label{sec/appendix}
\subsection{Inclination angles}\label{sec/stripe-theta}
We determined the instantaneous angles of stripes via the centred second moments $A_{ij}$ given by \cite{Tao2018extended}:
\begin{equation}
        A_{ij} = \frac{\int \mathrm{d}x \mathrm{d}z\, e x_i x_j}{\int \mathrm{d}x \mathrm{d}z\, e}
        - \left(\frac{\int \mathrm{d}x \mathrm{d}z\, e x_i}{\int \mathrm{d}x \mathrm{d}z\, e}\right)
          \left(\frac{\int \mathrm{d}x \mathrm{d}z\, e x_j}{\int \mathrm{d}x \mathrm{d}z\, e}\right)\,,
\end{equation}
where the indices $i,j$ both run over $x$ and $z$, with $x_x:=x$ and $x_z:=z$. In \cite{Tao2018extended} $e$ is ``\emph{defined as the disturbance kinetic energy relative to the base flow}''; in our work we set $e:=u_x^2(x,y^\star,z)$, with $y^\star=0$ for \pcf{} and $y^\star=0.46$ for \ppf{}. These are the wall-normal planes where the mean velocity profiles intersect the laminar profiles, that is, where the velocity fluctuations sum to zero. We truncate $e$ such that any $e(x,z)$ less than $0.005 \max [u_x^2(x,y^\star,z)]$ is set to zero.

One can then read off the angle $\theta$ using the expression
\begin{equation}
  \label{inst-angle}
    \theta = \frac{1}{2}\arctan\left(\frac{2A_{xz}}{A_{xx}-A_{zz}}\right)\,.
\end{equation}
We plotted the resulting angles at \figref{inst-angles}(a), along with their sines in \figref{inst-angles}.
A length $L$ can be found likewise from the expression
\begin{equation}
  \label{inst-length}
    \theta = \sqrt{12 (A_{xx} + A_{xz} \tan \theta)}.
\end{equation}

\begin{figure}
  \begin{overpic}[width=0.99\linewidth]{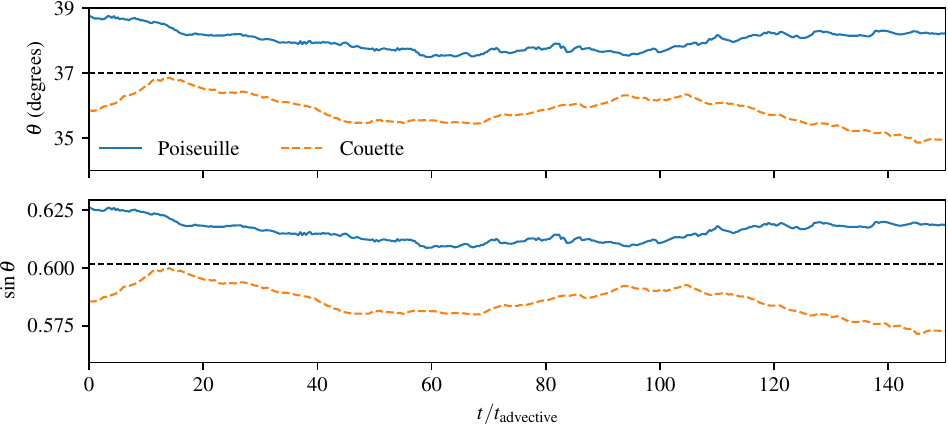}
    \put (-1,43) {(a)}
    \put (-1,25) {(b)}
  \end{overpic}
  \caption{(a) Instantaneous inclination angles \eqref{inst-angle} of \ppf{} ($\Rey^P=660$, solid blue) and \pcf{} ($\Rey^C=660$, dashed orange) and (b) the corresponding sines. We used fixed angles of $\theta^P=39^\degree$ and $\theta^C=37^\degree$ for \ppf{} and \pcf{}, respectively, as the inclination angles of the coordinates $x'-z'$ defined in \eqref{totilted} and shown in \figref{isolated-stripe}.}
  \label{fig/inst-angles}
\end{figure}

\subsection{Centres of mass}\label{sec/stripe-cm}
We compute the instantaneous centre of mass of stripes on the $x-z$ plane following \cite{Bai2009calculating}, by mapping each of the periodic coordinates $x$ and $z$ to a unit circle, taking $u_x^2(x, y^\star, z)$ as `mass' (with $y^\star$ as noted in \secref{stripe-theta}), computing the centre of mass within the corresponding circles, and inverting the resulting points back to $x$ and $z$ respectively. We can then find the $x'$ and $z'$ coordinates of the centre of mass using \eqref{totilted}.

More explicitly, to find the centre of mass in $x$, we first find the angles
\begin{equation}
  \theta_x = 2\pi x / L_x\,,
\end{equation}
which correspond to the points $(\tilde{x}_x, \tilde{z}_x)$ on a unit circle,
\begin{equation}
  \begin{aligned}
    \tilde{x}_x(\theta_x) &= \cos\theta_x\,,\\
    \tilde{y}_x(\theta_x) &= \sin\theta_x\,.
  \end{aligned}
\end{equation}
One can then find the centre of mass $(\bar{\tilde{x}}_x, \bar{\tilde{y}}_x)$ within this unit circle using an average weighted with $u_x^2(\theta_x, y^\star, z)$,
\begin{equation}
\begin{aligned}
  \bar{\tilde{x}}_x &= \sum_{xz} u_x^2(\theta_x, y^\star, z)\, \tilde{x}_x(\theta_x) / \sum_{xz} u_x^2(\theta_x, y^\star, z)\,,\\
  \bar{\tilde{y}}_x &= \sum_{xz} u_x^2(\theta_x, y^\star, z)\, \tilde{y}_x(\theta_x) / \sum_{xz} u_x^2(\theta_x, y^\star, z)\,.
\end{aligned}
\end{equation}
The angle of $(\bar{\tilde{x}}_x, \bar{\tilde{y}}_x)$,
\begin{equation}
  \bar{\theta}_x = \tan^{-1}(\bar{\tilde{y}}_x/\bar{\tilde{x}}_x)\,,
\end{equation}
gives the centre of mass in $x$,
\begin{equation}
  x_\mathrm{CM} = L_x \bar{\theta}_x / (2\pi)\,.
\end{equation}
The same algorithm is used for finding the centre of mass in $z$.

\FloatBarrier
\subsection{Additional data}
Here, we report additional data supporting our results of \secref{growth} and \secref{splitting}. \Figref{spacetime2} shows the spacetime plots of two additional trajectories, one per geometry, used in the analysis of internal streak creation (\secref{growth}). Figures \ref{fig/splitting-more-channel} and \ref{fig/splitting-more-couette} show two additional cases each of splitting events in \ppf{} (one case from the experiments of \cite{mukund2021aging} and one from our simulations) and \pcf{} (both from our simulations), respectively (see  \secref{splitting}).

% Spacetime plots
\begin{figure}
     \begin{overpic}[width=0.99\linewidth]{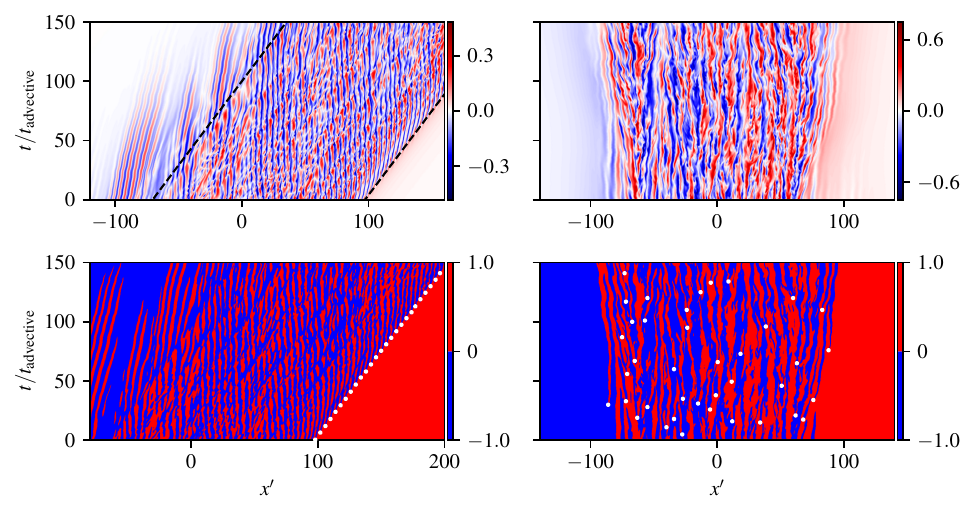}
         \put (0, 51)  {(a)}
         \put (51, 51)  {(c)}
         \put (0, 26)  {(b)}
         \put (51, 26)  {(d)}
     \end{overpic}
     \caption{(Same as \figref{spacetime}, but from different initial conditions.) Spacetime plots of streamwise velocity fluctuations in the bulk frames of reference of (a,b) \ppf{} and (c,d) \pcf{}. Colours are the values of (a) $u_x(t,x' + tU_\mathrm{bulk}\cos\theta^P)|_{y=0.46,\, z'=z'_\mathrm{CM}(t)+5}$ for \ppf{} and (c) $u_x(t,x')|_{y=0,\, z'=z'_\mathrm{CM}(t)}$ for \pcf{} (see dashed lines in panels (b) and (h) of \figref{isolated-stripe-cuts}). Panels (b,d) are binary versions of (a,c): anywhere with positive/negative streamwise velocity fluctuation is shown red/blue and newly created streaks are marked with white dots. For visualizations reasons, only the markings of positive (red) new streaks are shown for \ppf{}. Dashed black line downstream in (a) is fitted to the streak creation events marked in (b), the line upstream with the same slope is put as a guide.}
     \label{fig/spacetime2}
\end{figure}

\begin{figure}
    \begin{overpic}[width=0.325\linewidth]{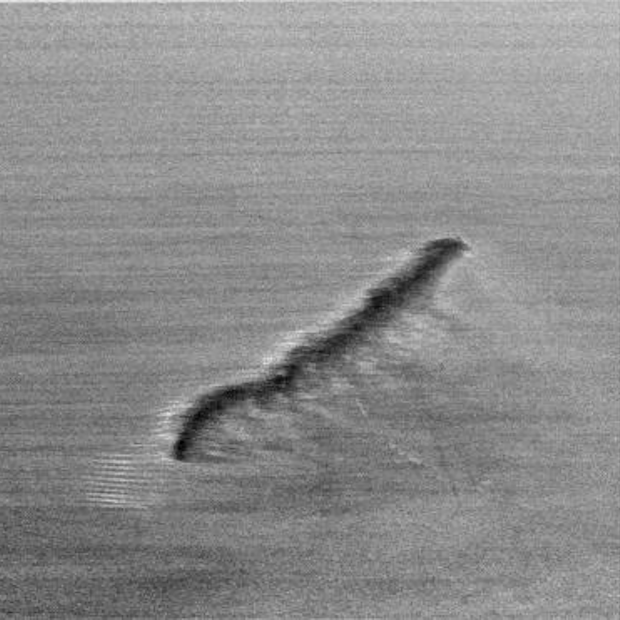}
        \put (-15,95) {(a)}
    \end{overpic}\,
    \begin{overpic}[width=0.325\linewidth]{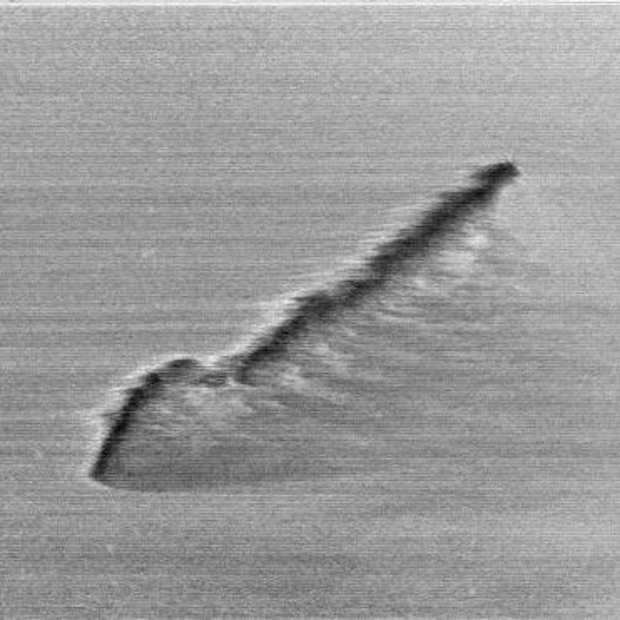}
        \put (50,102) {\large{Time $\rightarrow$}}
    \end{overpic}\,
    \begin{overpic}[width=0.325\linewidth]{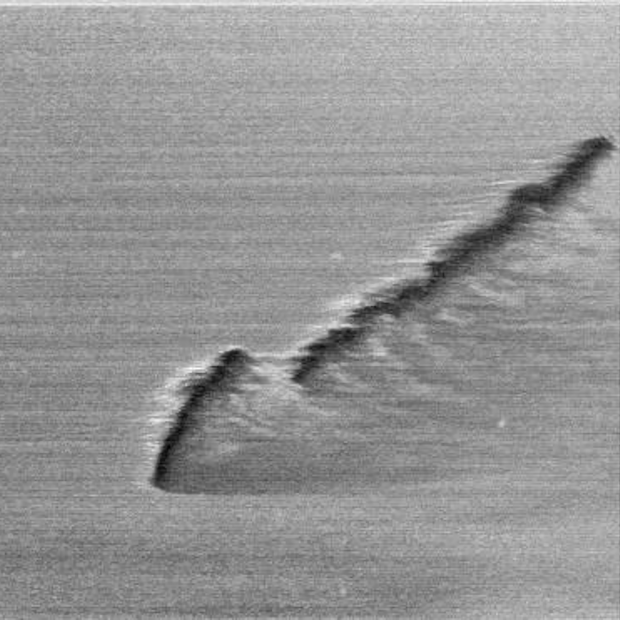}
    \end{overpic}\\
    \begin{overpic}[width=0.33\linewidth]{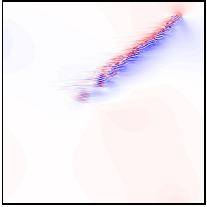}
        \put (-15,95) {(b)}
        \put (-25,100) {\rotatebox[origin=c]{90}{\large{$z \rightarrow$}}}
    \end{overpic}
    \begin{overpic}[width=0.33\linewidth]{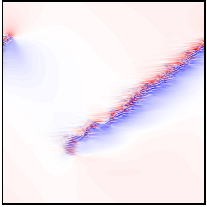}
        \put (50,-5) {\large{$x \rightarrow$}}
    \end{overpic}
    \begin{overpic}[width=0.33\linewidth]{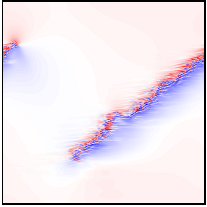}
    \end{overpic}\\
    \caption{Stripe evolution and proliferation in \ppf{} at $\Rey^P=750$ (a) from the experiments of \cite{mukund2021aging} and (b) from our simulations.}
    \label{fig/splitting-more-channel}
\end{figure}

\begin{figure}
    \begin{overpic}[width=0.246\linewidth]{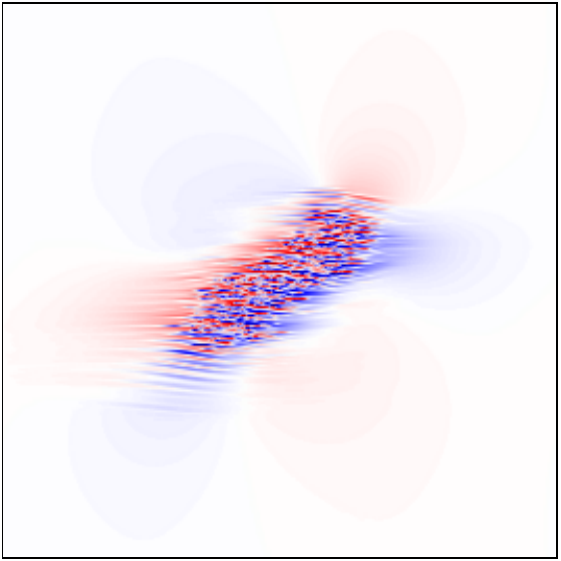}
        \put (-15,95) {(a)}
    \end{overpic}
    \begin{overpic}[width=0.246\linewidth]{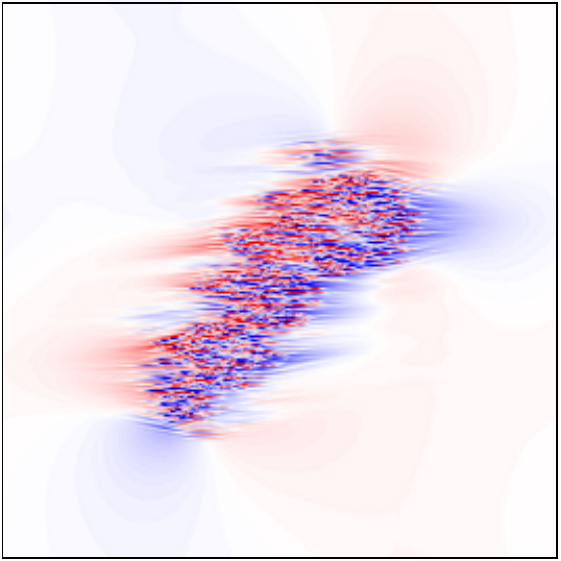}
        \put (85,102) {\large{Time $\rightarrow$}}
    \end{overpic}
    \begin{overpic}[width=0.246\linewidth]{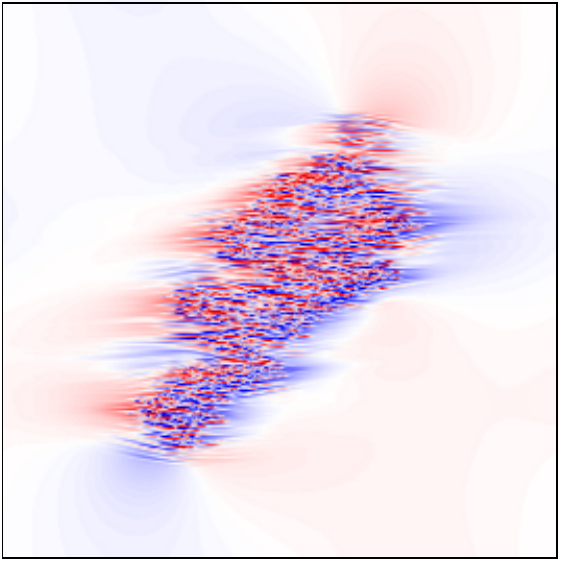}
    \end{overpic}
    \begin{overpic}[width=0.246\linewidth]{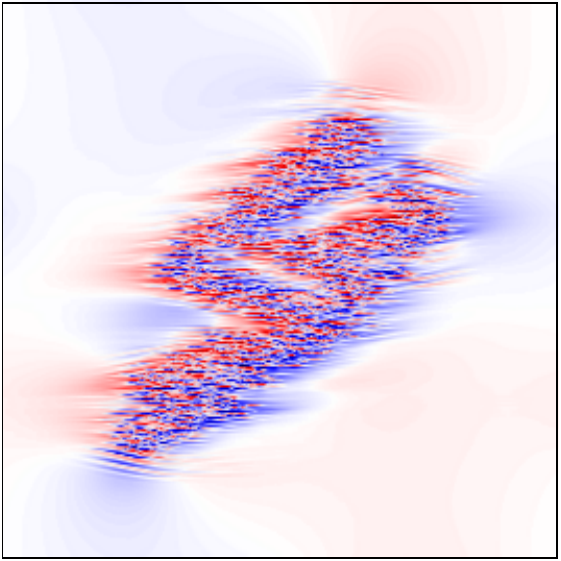}
    \end{overpic}\\
    \begin{overpic}[width=0.246\linewidth]{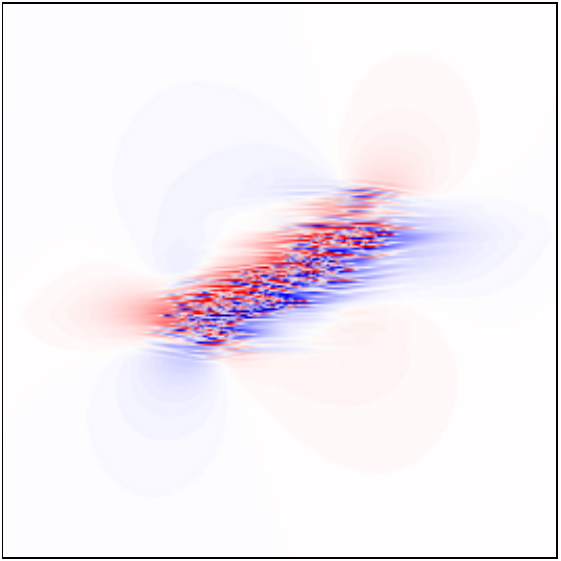}
        \put (-15,95) {(b)}
        \put (-25,100) {\rotatebox[origin=c]{90}{\large{$z \rightarrow$}}}
    \end{overpic}
    \begin{overpic}[width=0.246\linewidth]{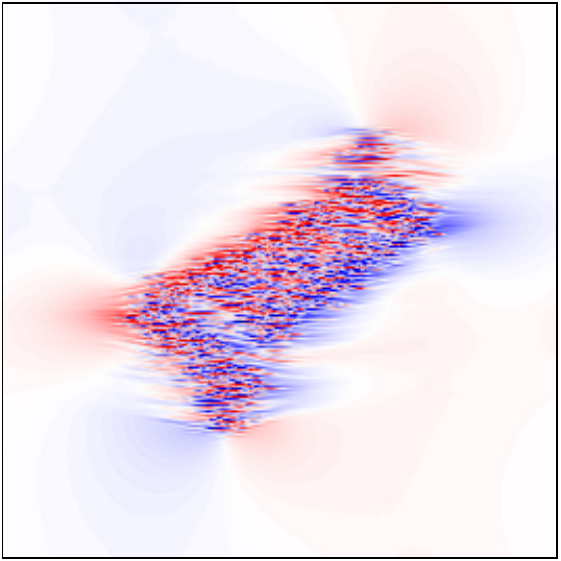}
        \put (95,-9) {\large{$x \rightarrow$}}
    \end{overpic}
    \begin{overpic}[width=0.246\linewidth]{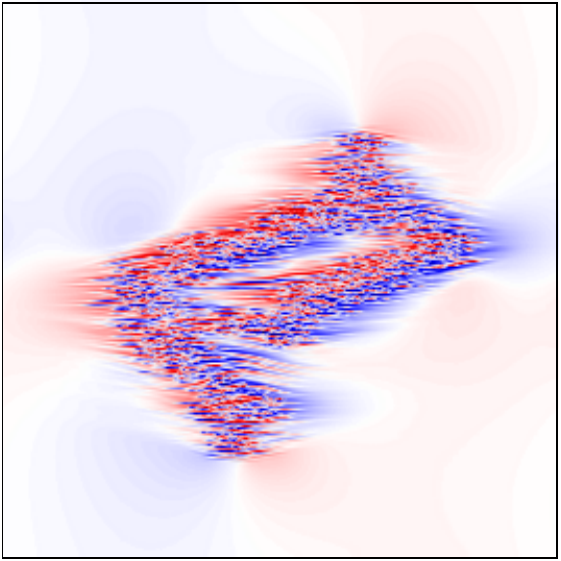}
    \end{overpic}
    \begin{overpic}[width=0.246\linewidth]{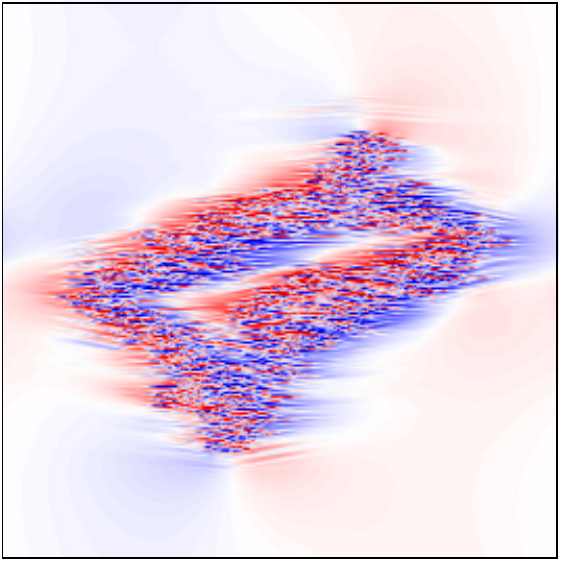}
    \end{overpic}
    \caption{Stripe evolution and proliferation in \pcf{} at $\Rey^C=350$. Streamwise velocity fluctuations $u_x$ are visualized in the $x$-$z$ plane at $y=0$.}
    \label{fig/splitting-more-couette}
\end{figure}

\printbibliography

\end{document}